\documentclass[%
reprint,
superscriptaddress,
amsmath,amssymb,
prb
]{revtex4-1}

\usepackage{amscd}
\usepackage[usenames,dvipsnames]{xcolor}
\usepackage{tensor}
\usepackage{braket,amsthm}
\usepackage{graphicx}
\usepackage{subcaption}
\usepackage{dcolumn}
\usepackage{bm}
\usepackage{hyperref}
\usepackage{epstopdf}

\def\epq{\epsilon_{\vec{q}}}
\DeclareMathOperator{\Tr}{Tr}
\DeclareMathOperator{\V}{\mathcal{V}}
\DeclareMathOperator{\bV}{\bar{\V}}

\makeatletter
\newcommand\footnoteref[1]{\protected@xdef\@thefnmark{\ref{#1}}\@footnotemark}
\makeatother

\begin{document}

\preprint{APS/123-QED}
\title{Density Matrix Modeling of Quantum Cascade Lasers without an Artificially Localized Basis:\\ 
	A Generalized Scattering Approach}
\author{Andrew Pan}
\email{Email: pandrew@ucla.edu.}
\affiliation{Department of Electrical Engineering, University of California, Los Angeles}
\author{Benjamin A. Burnett}
\affiliation{NG Next, Northrop Grumman Corporation, One Space Park, Redondo Beach, California}
\affiliation{Department of Electrical Engineering, University of California, Los Angeles}
\author{Chi On Chui}
\affiliation{Department of Electrical Engineering, University of California, Los Angeles}
\author{Benjamin S. Williams}
\affiliation{Department of Electrical Engineering, University of California, Los Angeles}
\date{\today}

\begin{abstract}
We derive a density matrix (DM) theory for quantum cascade lasers (QCLs) that describes the influence of scattering on coherences through a generalized scattering superoperator. The theory enables quantitative modeling of QCLs, including localization and tunneling effects, using the well-defined energy eigenstates rather than the ad hoc localized basis states required by most previous DM models. Our microscopic approach to scattering also eliminates the need for phenomenological transition or dephasing rates. We discuss the physical interpretation and numerical implementation of the theory, presenting sets of both energy-resolved and thermally averaged equations which can be used for detailed or compact device modeling. We illustrate the theory's applications by simulating a high performance resonant-phonon terahertz (THz) QCL design which cannot be easily or accurately modeled using conventional DM methods. We show that the theory's inclusion of coherences is crucial for describing localization and tunneling effects consistent with experiment.
\end{abstract}

\maketitle

\section{Introduction}
Quantum cascade lasers (QCLs) are important coherent light sources in the mid-infrared (MIR) and terahertz (THz) frequency regimes.\cite{faist_quantum_1994,yao_mid-infrared_2012,kohler_terahertz_2002,williams_terahertz_2007} Their versatility comes through heterostructure engineering of intersubband optical transitions; lasing occurs via an intricate balance of quantum tunneling, scattering (via disorder, phonons, electron-electron interactions, etc.), and optical coupling. A range of techniques has been developed to better understand and predict device operation.\cite{jirauschek_modeling_2014} Simple rate equations can explain basic features but are insufficient for quantitatively describing sophisticated QCL designs. At the other extreme, non-equilibrium Green's functions (NEGF) can provide detailed microscopic insight, but lead to considerable computational and physical complexity.\cite{lee_nonequilibrium_2002,schmielau_nonequilibrium_2009,kubis_theory_2009} Frequently, a balance of rigor and simplicity in modeling is sought using semiclassical or density matrix (DM) approaches, which may be solved analytically or numerically depending on the level of detail required. The advantages and limitations of these techniques are closely linked to the chosen basis of wave functions for the simulated device.

Semiclassical models use the eigenstates of the device Hamiltonian (i.e., the subbands generated by the heterostructure band structure and applied bias) as their basis with Fermi golden rule (FGR) scattering rates driving transitions between these states. They can be solved numerically as a set of self-consistent rate equations or via the Monte Carlo method, similar to the Boltzmann equation.\cite{chen_self-consistent_2011,jirauschek_monte_2010} These models can be also viewed as a type of Pauli master equation where only the diagonal elements (populations) of the density matrix are considered and off-diagonal elements (coherences) are neglected.\cite{kohn_quantum_1957,fischetti_theory_1998} While this approach has been very successful in describing many semiconductor device phenomena, its neglect of coherences can lead to problems in QCLs. For example, it is known that tunneling through injection barriers is not properly captured in semiclassical calculations,\cite{sirtori_resonant_1998,callebaut_importance_2005} which predict strong subthreshold parasitic current channels and a peak injection current density independent of barrier thickness (both of which are contradicted by experiment).

To compensate for these deficiencies, coherences between states must be taken into account. This is usually done phenomenologically in ``localized basis'' DM models by choosing a basis of wave functions localized on either side of particular barriers which couple coherently via tunneling matrix elements.\cite{sirtori_resonant_1998,kazarinov_possibility_1971}
This approach\cite{callebaut_importance_2005,kumar_coherence_2009,dupont_simplified_2010,dinh_extended_2012} is intuitively appealing and has clear physical significance in simple systems where a single thick barrier is the bottleneck for current. Therefore most self-described DM QCL models in the literature follow this approach. However, the results of such calculations are sensitive to the choice of basis, so they are not portable to different designs. Furthermore, the determination of the appropriate basis states and tunnel couplings is ad hoc and may be indefinable for complex designs where many states (and their effective couplings) must be disentangled. This is particularly troublesome in many THz devices, where the low energy scale necessitates many closely spaced states in energy and position and resonant tunneling is critical for depopulation transport within the module--it is not always easy to decide in advance that particular barriers are bottlenecks. It can also be important in MIR QCLs, where coherent tunneling interactions mediate the miniband extraction process. Finally, the juxtaposition of an artificially localized basis with FGR scattering rates calculated from energy eigenstates is theoretically unsatisfactory. It is therefore highly desirable to devise a theory which consistently accounts for quantum coherences without any ambiguity regarding the choice of basis.

Ideally the energy eigenstates, which are easily computed using band structure solvers, would be used to describe coherences. Since the Hamiltonian is diagonal, only scattering can then induce off-diagonal DM elements. An early QCL simulation work explored this idea; however, few details of the formalism were given and the authors concluded that coherences were unimportant in steady state for the device they considered.\cite{iotti_nature_2001} As noted above, strong evidence is now known for coherence effects in QCLs. More recently, a few works have described QCL DM models in the time domain including scattering-induced coherences.\cite{weber_density-matrix_2009,jonasson_partially_2016,jonasson_quantum_2016} Ref. \onlinecite{weber_density-matrix_2009} used this method to show the importance of coherences for tunneling transport and dynamical charge transfer, while Refs. \onlinecite{jonasson_partially_2016,jonasson_quantum_2016} examined nonequilibrium populations and relaxation times of QCL subbands. However, an extensive unified discussion of the theory, its inner workings, and its applications has not yet appeared.

In this paper, we present such a description of a DM theory derived from first principles which fully captures coherence effects within the energy eigenstate basis. In this theory, a generalized scattering superoperator appears which can not only redistribute populations but also induce and dephase coherences, allowing for transfer between arbitrary DM elements. From the implementation standpoint, this amounts to an extension of the FGR to include effects of scattering on off-diagonal DM elements. Relevant quantities like charge density, current, gain, etc., can be then be computed using standard methods. Because the model uses the energy eigenstates of the QCL module and does not require any phenomenological parameters to describe dephasing, tunneling, or other effects, it can be directly applied to different QCL designs without modification. As an initial example of its capabilities, we apply our theory to a five-level THz QCL design not amenable to existing DM methods, finding good agreement with experiment and insight into device operation. Aside from certain technical differences in our derivation and equations which are addressed below, this work differs from related previous studies\cite{weber_density-matrix_2009,jonasson_partially_2016,jonasson_quantum_2016} by providing a complete framework and physical discussion of steady-state and optical modeling of QCLs, and we further point out that this approach resolves the basis choice dilemma of conventional QCL DM modeling.

As this paper seeks to cover our approach from fundamental derivation to practical implementation, the various aspects of the theory will be discussed as follows.
Section \ref{sec:derivation} explains the derivation of the DM equations, showing how a generalized scattering superoperator emerges from the Liouville-von Neumann equation for the microscopic density matrix of a general system. Section \ref{sec:features} discusses the physical interpretation of the superoperator and its relationship with other scattering models. In Section \ref{sec:scattering} we provide equations for generalized scattering via impurities and alloy disorder, interface roughness, and polar optical phonons, discussing how these mechanisms can be included in both microscopic (energy-resolved) and coarse-grained (thermally averaged) form for detailed and simplified DM models, respectively. The incorporation of periodicity and optical field in the model is treated in Section \ref{sec:period}. In Section \ref{sec:examples} we summarize how the model can be used for device calculations (readers primarily interested in implementing this model may wish to start here) and then show numerical examples of its use for a two-level superlattice as well as a complex five-level THz QCL design. The device analysis in this paper emphasizes the interpretation of the formalism; we will discuss a wider range of QCL designs, as well as device-oriented insights obtained from our methodology, in a separate work.\cite{burnett_notitle_nodate}

\section{Generalized Scattering Theory}\label{sec:derivation}
To overcome the limitations of conventional modeling approaches, we examine the evolution of the density matrix in detail to determine how coherences arise microscopically within the energy eigenstate basis. Our method of derivation follows that of Luttinger and Kohn,\cite{kohn_quantum_1957,jones_theoretical_1973,fischetti_theory_1998} generalizing their result to include off-diagonal DM elements. For specificity, we consider the case of a coupled electron-phonon system within the Hilbert product space $\mathcal{H}_s = \mathcal{H} \otimes \mathcal{H}_{ph}$ of the electrons and phonons. This allows us to obtain a generalized model of the electronic DM in the presence of phonon scattering; the case of elastic scattering can be derived in similar if slightly simpler fashion, with results discussed later in the section.

We adapt the Liouville-von Neumann equation for the DM evolution of the complete system $\rho_s$ to write
\begin{equation}
\dot{\rho}_{s} \equiv \frac{\partial \rho_{s}}{\partial t} = \frac{1}{i\hbar}[H' + \mathcal{V},\rho_s] - \frac{\eta}{\hbar}\rho_s ,
\end{equation}
where we have separated the isolated Hamiltonians for the electrons and phonons $H' = H + H_{ph}$ from the electron-phonon coupling $\mathcal{V}$ and then formally introduce dissipation via an infinitesimal damping constant $\eta$ in the right-most term.\footnote{See \cite{kohn_quantum_1957}. Strictly speaking this term should be $\eta(\rho_s - \rho_{s,eq})$ where $\rho_{s,eq}$ is the equilibrium DM, but the latter term can be neglected since we will take the limit that $\eta \rightarrow 0$.} Since QCLs typically operate at nondegenerate electron densities, we neglect Pauli exclusion effects and assume a single-particle electron Hamiltonian $H$ which includes the band structure and any applied electric field (as well as the self-consistent Hartree potential if space charge effects are considered). Typically we work in the basis of eigenstates of $H'$ which have energies $E' = E + \sum\limits_{\vec{q}} (n_{\vec{q}} + 1/2) \epq$, where $E$ is the electron eigenenergy and $n_{\vec{q}}$ and $\epq$ are the occupation number and energy for each phonon mode $\vec{q}$ (shorthand for the set of quantum numbers denoting each mode including wave vector, dispersion branch, and polarization). For convenience, any diagonal components of $\mathcal{V}$ are also lumped into $E'$. The evolution for an arbitrary element $\rho_{s,ab}$ is then given by
\begin{equation}\label{eq:dm_dt1}
\begin{split}
\dot{\rho}_{s,ab} = \left(\frac{{E}'_{a} - E'_{b}}{i\hbar} - \frac{\eta}{\hbar} \right)\rho_{s,ab} + \sum\limits_{c} \frac{[\mathcal{V}_{ac}\rho_{s,cb} - \mathcal{V}_{cb}\rho_{s,ac}]}{i\hbar}
\end{split}
\end{equation}
If we assume that the density matrix varies slowly on the time scale of $H'$ and $\mathcal{V}$, we can drop the time derivative of $\rho_s$ on the left-hand side (akin to the Markov approximation) and rearrange to get
\begin{equation}\label{eq:dm_coh1}
\rho_{s,ab} = \frac{-1}{E'_a - E'_b - i\eta}\left[ \sum\limits_{c}\left(\mathcal{V}_{ac} \rho_{s,cb} - \mathcal{V}_{cb} \rho_{s,ac}\right) \right] .
\end{equation}
Anticipating integrals to come, we now make use of
\begin{equation}
\lim\limits_{\eta \rightarrow 0} \frac{1}{\omega - i\eta} = \mathcal{P}\left(\frac{1}{\omega}\right) + i\pi \delta(\omega)
\end{equation}
and drop the principal value contributions (describing renormalization of the energy levels) to obtain
\begin{equation}\label{eq:dm_coh2}
\rho_{s,ab} = -i\pi \left[ \sum\limits_{c}\left(\mathcal{V}_{ac} \rho_{s,cb} - \mathcal{V}_{cb} \rho_{s,ac}\right) \right]\delta(E'_a - E'_b) .
\end{equation}
By substituting Eq. \ref{eq:dm_coh2} back into the right-hand side of Eq. \ref{eq:dm_dt1} and slightly rearranging the indices, we find to second order in the scattering potential $\mathcal{V}$ that
\begin{equation}\label{eq:dm2ndorder1}
\begin{split}
& \dot{\rho}_{s,ab} = \frac{(E_a'-E_b')}{i\hbar}\rho_{s,ab} \\
&+ \frac{\pi}{\hbar} \sum\limits_{c,d} \left[\mathcal{V}_{ac}\mathcal{V}_{db}\rho_{s,cd}(\delta(E'_c - E'_b) + \delta(E'_a - E'_d))  \right.\\
&- \left. 
\mathcal{V}_{ac}\mathcal{V}_{cd}\rho_{s,db} \delta(E'_c - E'_b) - \mathcal{V}_{dc}\mathcal{V}_{cb}\rho_{s,ad}\delta(E'_a - E'_c) \right] .
\end{split}
\end{equation}
The terms within the double summation can be interpreted as a generalized scattering superoperator which couple the evolution of an arbitrary DM element $\rho_{s,ab}$ to all others. This is in contrast to the usual Fermi golden rule rates which only couple diagonal elements (populations).

For inelastic scattering via phonons, the electron-phonon interaction $\mathcal{V} = \sum\limits_{\vec{q}} V^{-\vec{q}} \hat{b}_{\vec{q}}^{\dag} + V^{\vec{q}}\hat{b}_{\vec{q}}$, where $\vec{q}$ denotes the phonon mode wave vector (with associated energy $\epq$), $V^{\vec{q}}$ is the electron-phonon matrix element for the given mode, and $\hat{b}_{\vec{q}}$ and $\hat{b}_{\vec{q}}^{\dag}$ are the phonon creation and annihilation operators, respectively. The products of matrix elements in Eq. \ref{eq:dm2ndorder1} have the form
\begin{equation}
\begin{split}
\mathcal{V}_{ac}\mathcal{V}_{db} = \sum\limits_{\vec{q}, \vec{q}'} & \left( V_{ac}^{-\vec{q}}V_{db}^{-\vec{q}'} \hat{b}_{\vec{q}}^{\dag}\hat{b}_{\vec{q}'}^{\dag} + V_{ac}^{-\vec{q}}V_{db}^{\vec{q}'} \hat{b}_{\vec{q}}^{\dag}\hat{b}_{\vec{q}'} \right. \\
& \left. + V_{ac}^{\vec{q}}V_{db}^{-\vec{q}'} \hat{b}_{\vec{q}}\hat{b}_{\vec{q}'}^{\dag} +
V_{ac}^{\vec{q}}V_{db}^{\vec{q}'} \hat{b}_{\vec{q}}\hat{b}_{\vec{q}'} \right)
\end{split}
\end{equation}
and likewise for other index orderings. Since we are interested in the electrons, we can perform a partial trace over the phonon degrees of freedom to obtain a master equation for the electron density matrix. We now assume that the electrons and phonons are weakly coupled so $\rho_s = \rho \otimes \rho_{ph}$ can be factorized into a tensor product of electron and phonon density matrices and that the phonon subsystem $\rho_{ph}$ is in equilibrium, so the population of each mode $n_{\vec{q}}$ is given by the Bose-Einstein occupation factor at the lattice temperature. Only terms like $\braket{\hat{b}_{\vec{q}}^{\dag}\hat{b}_{\vec{q}}} = n_{\vec{q}}$ and $\braket{\hat{b}_{\vec{q}}\hat{b}_{\vec{q}}^{\dag}} = n_{\vec{q}}+1$ survive the ensemble averaging, which correspond to phonon absorption and emission, respectively. Eq. \ref{eq:dm2ndorder1} therefore reduces to
\begin{equation}\label{eq:dm2ndorder1_electron}
\begin{split}
\dot{\rho}_{ab} = &\frac{(E_a -E_b)}{i\hbar}\rho_{ab} \\
+ \frac{\pi}{\hbar} \sum\limits_{c,d,\vec{q},\pm}& \left(n_{\vec{q}}+\frac{1}{2} \pm \frac{1}{2}\right) \left[V_{ac}^{\mp q}V_{db}^{\pm q}\rho_{cd}(\delta(E_c - E_b \mp \epq)  \right. \\ 
&+ \delta(E_a - E_d \pm \epq))  - 
V_{ac}^{\mp q}V_{cd}^{\pm q}\rho_{db} \delta(E_c - E_b \pm \epq) \\
&\left. - V_{dc}^{\mp q}V_{cb}^{\pm q}\rho_{ad}\delta(E_a - E_c \mp \epq)  \right] .
\end{split}
\end{equation}
An alternative notation useful for symbolic matrix manipulation of the superoperator is given in Appendix \ref{app:matnote}.

Thus far we have defined the DM using general labels $a, b \dots$ for the electronic basis. In quantum well-based QCLs, the basis states may depend on subband, transverse momentum, and periodicity, which leads to a complicated indexing of operator and superoperator elements. We will label subband indices (denoting the subband envelope functions) with upper case Roman subscripts and the in-plane (transverse) momentum $\vec{k}=(k_x,k_y)$ with lower case Roman superscripts, while module periods (introduced later in Section \ref{sec:period}) will be denoted with Greek subscripts $\mu$. 
Since the device Hamiltonian $H$ is usually diagonal in transverse momentum $\vec{k}$, we can rewrite the evolution of an arbitrary element of the electron DM Eq. \ref{eq:dm2ndorder1_electron} as
\begin{equation}\label{eq:DM_label}
\dot{\rho}_{AB}^{\vec{k}} = \frac{i}{\hbar}(E_B^{\vec{k}} - E_A^{\vec{k}})\rho_{AB}^{\vec{k}} + \sum\limits_{C,D,\vec{k}'}\Gamma^{\vec{k},\vec{k}'}_{AB,CD}\rho_{CD}^{\vec{k}'}
\end{equation}
where the generalized scattering superoperator $\Gamma$ describing scattering from the DM element $[CD,\vec{k}']$ to $[AB,\vec{k}]$ is given by
\begin{equation}\label{eq:gen_so}
\begin{split}
\Gamma_{AB,CD}^{\vec{k},\vec{k}'} &= \frac{\pi}{\hbar} \sum\limits_{\vec{q},\pm} \left(n_{\vec{q}} + \frac{1}{2} \pm \frac{1}{2}\right) \left[ V_{A,C}^{\vec{k},\vec{k}'} V_{D,B}^{\vec{k}',\vec{k}}  \times \right. \\
& \left(\delta(E_A^{\vec{k}}-E_D^{\vec{k}'} \pm \epsilon_{\vec{q}}) + \delta(E_B^{\vec{k}}-E_C^{\vec{k}'} \pm \epsilon_{\vec{q}}) \right) \\
& - \delta_{\vec{k},\vec{k}'}\sum\limits_{F,\vec{k}''} \left( \delta_{A,C} V^{\vec{k},\vec{k}''}_{D,F}V^{\vec{k}'',\vec{k}}_{F,B} \delta(E_F^{\vec{k}''}-E_A^{\vec{k}} \pm \epsilon_{\vec{q}}) \right. \\
&\left. \left. + \delta_{B,D} V^{\vec{k},\vec{k}''}_{A,F}V^{\vec{k}'',\vec{k}}_{F,C} \delta(E_F^{\vec{k}''}-E_B^{\vec{k}} \pm \epsilon_{\vec{q}}) \right) \right]
\end{split}
\end{equation}
for inelastic scattering. In this equation the Kronecker delta function $\delta_{i,j}$ is 1 if $i=j$ and zero otherwise. For elastic scattering mechanisms $V$ we can follow a similar derivation for the electron DM to obtain the associated superoperator
\begin{equation}\label{eq:gen_so_elastic}
\begin{split}
\Gamma_{AB,CD}^{\vec{k},\vec{k}'} &= \frac{\pi}{\hbar} \left[ V_{A,C}^{\vec{k},\vec{k}'} V_{D,B}^{\vec{k}',\vec{k}}  \left(\delta(E_A^{\vec{k}}-E_D^{\vec{k}'}) + \delta(E_B^{\vec{k}}-E_C^{\vec{k}'}) \right) \right. \\
& - \delta_{\vec{k},\vec{k}'}\sum\limits_{F,\vec{k}''} \left( \delta_{A,C} V^{\vec{k},\vec{k}''}_{D,F}V^{\vec{k}'',\vec{k}}_{F,B} \delta(E_F^{\vec{k}''}-E_A^{\vec{k}}) \right. \\
&\left. \left. + \delta_{B,D} V^{\vec{k},\vec{k}''}_{A,F}V^{\vec{k}'',\vec{k}}_{F,C} \delta(E_F^{\vec{k}''}-E_B^{\vec{k}}) \right) \right] .
\end{split}
\end{equation}
In these equations we ignore coherences between different $\vec{k}$ vectors, i.e., we assume that $\rho_{AB}^{\vec{k}_1 \vec{k}_2}$ and $\Gamma_{AB,CD}^{\vec{k}_1\vec{k}_2,\vec{k}_3\vec{k}_4}$ are zero when $\vec{k}_1 \neq \vec{k}_2$ and/or $\vec{k}_3 \neq \vec{k}_4$, which is justified by the in-plane translational invariance of the quantum wells in QCLs. This ignores the possibility of in-plane localization due to strong disorder;\cite{ndebeka-bandou_importance_2013} however such effects arise from multiple scattering and are beyond the scope of the FGR or other low-order scattering treatments in any case.

Eqs. \ref{eq:DM_label}-\ref{eq:gen_so_elastic} are the basic results of this paper and suggest a different picture from the usual semiclassical or phenomenological DM theories. In contrast to phenomenological methods, this model does not require ad hoc determination of a localized basis, instead working with the well-defined and easily calculated energy eigenstates of the system. Furthermore, the generalized scattering superoperator Eq. \ref{eq:gen_so} not only provides for transitions between different populations (as in the semiclassical theory) and dephasing of coherences, but also couples arbitrary populations and coherences with each other. Note that the sign of these terms may be positive or negative, depending on the phases of the wave functions involved in the scattering process; as we discuss below, this can be connected with the physical localization of charge in a particular spatial region.

The neglect of the time derivative in Eq. \ref{eq:dm_coh1} is justifiable in the steady state DC limit, assuming it exists for the system. Under transient conditions or with a time-dependent excitation such as an optical field, the validity of the assumption depends on the relative time/energy scales of the band structure, excitations, and scattering mechanisms. Non-Markovian effects may need to be considered when these scales become intertwined.\cite{breuer_theory_2002,weiss_quantum_2008} However, the success of our model in describing a variety of designs and device phenomena suggests that this remains a good approximation for most practical QCLs.

\section{Features of the Theory}\label{sec:features}
\subsection{Relationship with Simple DM Models}\label{sec:relationship}
\begin{figure}[htb]
	\includegraphics[width=4.5cm]{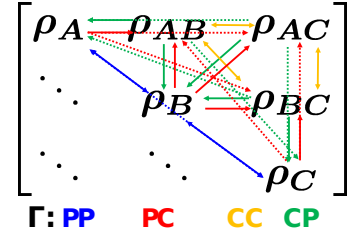}
	\caption{Categories of superoperator terms (denoted by arrows) and the coupling they induce between elements of the density matrix. For clarity the transverse momentum $\vec{k}$, $\vec{k}'$ is suppressed in the illustration. See text for discussion.}\label{fig:superop_map}
\end{figure}
To obtain a clearer physical picture of the theory, we can write the evolution for an arbitrary DM population or coherence element
\begin{align}\label{eq:dm_resolved}
\dot{\rho}_A^{\vec{k}} = &\sum\limits_{C,\vec{k}'}\Gamma_{A,C}^{\vec{k},\vec{k}'}\rho_C^{\vec{k}'}
+ \sum\limits_{C \neq D, \vec{k}'}\Gamma_{A,CD}^{\vec{k},\vec{k}'}\rho_{CD}^{\vec{k}'} \\
\dot{\rho}_{AB}^{\vec{k}} = &\frac{i}{\hbar}[E_B^{\vec{k}} - E_A^{\vec{k}}]\rho_{AB}^{\vec{k}} +\sum\limits_{C,\vec{k}'}\Gamma_{AB,C}^{\vec{k},\vec{k}'}\rho_{C}^{\vec{k}'} \\ \nonumber
&+ \sum\limits_{C\neq D,\vec{k}'}\Gamma_{AB,CD}^{\vec{k},\vec{k}'}\rho_{CD}^{\vec{k}'}
\end{align}
where for brevity we label populations (diagonal DM elements) $AA = A$, etc. Note that the delta functions within the $\Gamma$ superoperator (Eqs. \ref{eq:gen_so}-\ref{eq:gen_so_elastic}) specify the values of $\vec{k}'$ which contribute within each summation. As illustrated in Fig. \ref{fig:superop_map}, we see that the generalized scattering superoperator yields several distinct types of terms, describing scattering from population to population (PP) $\Gamma_{A,C}^{\vec{k},\vec{k}'}$, coherence to population (CP) $\Gamma_{A,CD}^{\vec{k},\vec{k}'}$, population to coherence (PC) $\Gamma_{AB,C}^{\vec{k},\vec{k}'}$, and coherence to coherence (CC) $\Gamma_{AB,CD}^{\vec{k},\vec{k}'}$. The PP terms describe transitions between the subband states and reduce to the FGR rate
\begin{equation}\label{eq:fgr}
\Gamma_{A,C}^{\vec{k},\vec{k}'} = \frac{2\pi}{\hbar}|V_{A,C}^{\vec{k},\vec{k}'}|^2\delta(E_A^{\vec{k}} - E_C^{\vec{k}'})
\end{equation}
for elastic scattering (and similarly for inelastic processes) when $A \neq C$. 
Likewise, CC self-couplings (i.e., $\Gamma_{AB,AB}^{\vec{k},\vec{k}}$) are equivalent to coherence dephasing rates. In the terminology of Bloch equations, PP rates and CC self-couplings are equivalent to the $T_1$ and $T_2$ relaxation times for populations and coherences.\cite{iotti_quantum_2005} The other categories of terms (CP, PC, and general CC couplings such as $\Gamma_{AB,CD}^{\vec{k},\vec{k}'}$) do not usually appear in phenomenological DM theories but are needed to describe how coherences evolve through scattering. This distinguishes the present theory from semiclassical models which only retain PP terms and localized models where coherences are built up through a specific choice of the coherent Hamiltonian basis (e.g., off-diagonal tunnel couplings in $H$).

Because we define all scattering terms using a single superoperator, the sign of each $\Gamma_{AB,CD}^{\vec{k},\vec{k}'}$ term depends on whether the process it describes is contributing to the buildup or decay of the DM element $[AB, \vec{k}]$ from $[CD, \vec{k}']$. It is clear that the FGR $(\Gamma_{A,C}^{\vec{k},\vec{k}'})$ rates are always positive since they describe transfer of population into the final state. Similarly, the population inverse lifetimes $\Gamma_{A,A}^{\vec{k},\vec{k}}$ and dephasing rates $\Gamma_{AB,AB}^{\vec{k},\vec{k}}$ always have negative signs in our formulation because they describe decay or transfer out of the DM element. However, we can see from the scattering superoperator Eq. \ref{eq:gen_so} that the signs of all other terms (PC, CP, CC) can be either positive or negative as they do not necessarily involve conjugate products of the scattering potential $V$. Nonetheless the scattering superoperator is always real for physical processes, as will be clear when we derive $\Gamma$ for specific scattering mechanisms.
\begin{figure}[htb]
	\includegraphics[width=8.8cm]{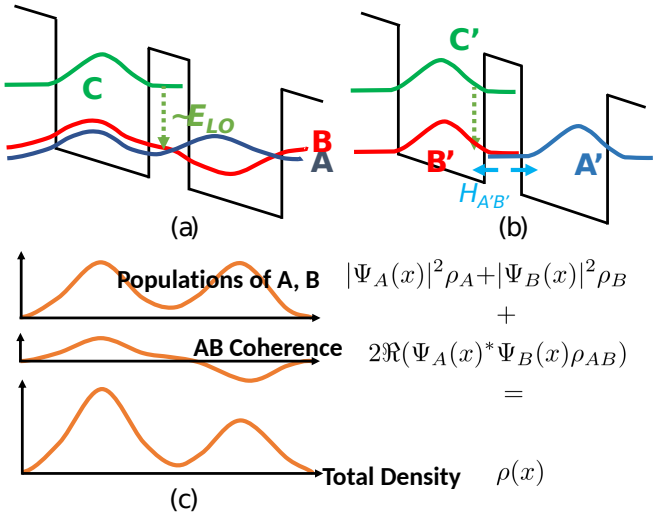}
	\caption{Wave functions of a three-level system with (a) energy eigenstates and (b) ad hoc localized states. (c) Contributions to electron density from populations and coherence of anticrossed energy states $A$ and $B$; positive $\rho_{AB}$ leads to charge buildup in the ``upstream'' well to the left (as pictured), and vice versa. The transverse momentum $\vec{k}$ dependence is suppressed for clarity.}\label{fig:threelevelstates}
\end{figure}

We can obtain more physical intuition for the different scattering terms  by noting that for the spatially extended energy eigenstates of QCLs, coherences can be partially interpreted as spatial localization of electrons. Scattering ``reshapes'' the occupied states by building up coherences between them. Consider the two anticrossed extended subbands $A$ and $B$ depicted in Fig. \ref{fig:threelevelstates}a for instance. In phenomenological theories, this is described using an artificial basis of states localized on either side of the barrier $A'$ and $B'$, which couple to each other via a tunneling matrix element $H_{A'B'}$, as pictured in Fig. \ref{fig:threelevelstates}b. In the energy eigenstate basis that this theory uses, the spatial charge distribution is obtained from the density matrix as shown in Fig. \ref{fig:threelevelstates}c, so the magnitude and sign of coherences between the subbands $\rho_{AB}^{\vec{k}}$ describes charge buildup on either side of the barrier.

From this perspective a PC transition, for instance from a higher energy state $C$ into the coherence of $AB$, describes scattering into localized regions in either the left or right well, which are distinguished by the sign of the superoperator term. For the system pictured in Fig. \ref{fig:threelevelstates}a, we can see from Eq. \ref{eq:gen_so} that $\Gamma_{AB,C}^{\vec{k},\vec{k}'} \propto \sum\limits_{\vec{q}}V_{A,C}^{\vec{k},\vec{k}'}V_{C,B}^{\vec{k'},\vec{k}}$, which may be positive or negative depending on the matrix elements. It is instructive to consider the example of long wavelength phonon scattering where $V_{AB}$ is simply proportional to the wave function overlap; in this case, $\Gamma_{AB,C}^{\vec{k},\vec{k}'}$ is positive for the polarity of the wave functions in Fig. \ref{fig:threelevelstates}a, implying charge localization in the left well behind the tunnel barrier. (If the phases of the wave functions were chosen differently, so that for instance the sign of state $B$ is flipped, that will change the sign of $\Gamma_{AB,C}^{\vec{k},\vec{k}'}$ and $\rho_{AB}$ without any physical consequences, since the contribution to the charge density will remain in the upstream well.) Similarly, a CP transition (say from $AB$ to $A$ or $B$) might drive delocalization, CC terms describes scattering between localized regions, etc. It is clear that the different categories of superoperator couplings account for different spatial transfers of charge density compared to the FGR/PP coupling between extended eigenstates.

\subsection{Population Conservation and Sum Rules}
It is well known that the FGR transition rates or PP terms Eq. \ref{eq:fgr} satisfy the detailed balance ``sum rule'' $\sum\limits_{A,\vec{k},\vec{k}'} \Gamma_{A,C}^{\vec{k},\vec{k}'} = 0$, which is necessary to preserve the DM trace and conserve particle number. Another useful sum rule for the generalized superoperator applies to CP coupling, which describes ``transfer'' of electrons from coherences into populations. From Eq. \ref{eq:gen_so}, we see that for a generic CP term, if we sum over all possible $\vec{k}'$ and final $\vec{k}$ and $A$, we obtain
\begin{equation}
\begin{split}
\sum\limits_{A,\vec{k},\vec{k}'} \Gamma_{A,CD}^{\vec{k},\vec{k}'} = &\frac{\pi}{\hbar} \sum\limits_{\vec{k}, \vec{k}',\vec{q},\pm} \left(n_{\vec{q}} + \frac{1}{2} \pm \frac{1}{2}\right) \left[  \sum\limits_A V_{A,C}^{\vec{k},\vec{k}'} V_{D,A}^{\vec{k}',\vec{k}}  \times \right. \\
& \left(\delta(E_A^k-E_D^{k'} \pm \epsilon_{\vec{q}}) + \delta(E_A^{k}-E_C^{k'} \pm \epsilon_{\vec{q}}) \right) \\
& - \sum\limits_{F} \left( V^{\vec{k},\vec{k}'}_{D,F}V^{\vec{k}',\vec{k}}_{F,C} \delta(E_F^{k'}-E_C^{k} \pm \epsilon_{\vec{q}}) \right. \\
&\left. \left. + V^{\vec{k},\vec{k}'}_{D,F}V^{\vec{k}',\vec{k}}_{F,C} \delta(E_F^{k'}-E_D^{k} \pm \epsilon_{\vec{q}}) \right) \right] .
\end{split}.
\end{equation}
Switching $\vec{k}$ and $\vec{k}'$ in the summation over $F$, we find that the sums over $A$ and $F$ cancel and therefore obtain a simple sum rule for population-coherence couplings:
\begin{equation}
\sum\limits_{A,\vec{k},\vec{k}'} \Gamma_{A,CD}^{\vec{k},\vec{k}'} = 0
\end{equation}
The same relation evidently applies for elastic scattering by taking $\epq=0$ and setting $\sum\limits_{\vec{q}, \pm}(n_{\vec{q}} + \frac{1}{2} \pm \frac{1}{2})=1$. This implies that the net transfer of population through any DM coherence $CD$ must be zero, i.e., any portion of a population $\rho_A^{\vec{k}'}$ that ``scatters'' into a coherence must ultimately be redistributed to some other population $\rho_B^{\vec{k}}$. As with the FGR sum rule, this is necessary to preserve the trace and conserve population\cite{iotti_quantum_2005} and thus generalizes for any system, not just the quantum well-based structures we discuss in this paper. For device calculations, these properties may be practically useful for checking the consistency of calculations and inferring superoperator rates for simple systems.

\subsection{Schrodinger versus Interaction Pictures}\label{subsec:schrodint}
Formally, the theory presented here is a type of Redfield equation, being a Born-Markov master equation for the density matrix. \cite{redfield_theory_1957,weiss_quantum_2008} Such equations are in general not of the Lindblad form\cite{lindblad_generators_1976} and hence not guaranteed to be completely positive, so that it is possible to obtain negative populations, for instance for strong scattering or certain initial conditions.\cite{iotti_quantum_2005,weiss_quantum_2008} In practice, this treatment seems applicable to conventional QCLs, as we have used it for calculations in a wide variety of device designs and have rarely observed negative values, which generally only occur for states whose populations are orders of magnitude smaller than those of other states and thus have no noticeable impact on device properties.

The derivation presented above does differ in one significant way from similar formulations for semiconductor devices where the Liouville-von Neumann equation is written in integrodifferential form and then solved under the Born and Markov approximations in the interaction picture (where the unperturbed Hamiltonian $H'$ is absorbed into the time dependence of operators).\cite{iotti_quantum_2005,jonasson_partially_2016} The final results are very similar, but the interaction approach leads to a slightly different form of the scattering superoperator
\begin{equation}\label{eq:gen_so_interact}
\begin{split}
\Gamma_{AB,CD\text{ int}}^{\vec{k},\vec{k}'} &= \frac{\pi}{\hbar} \sum\limits_{\vec{q},\pm} \left(n_{\vec{q}} + \frac{1}{2} \pm \frac{1}{2}\right) \left[ V_{A,C}^{\vec{k},\vec{k}'} V_{D,B}^{\vec{k}',\vec{k}}  \times \right. \\
& \left(\delta(E_A^k-E_C^{k'} \pm \epq) + \delta(E_B^{k}-E_D^{k'} \pm \epq) \right) \\
& - \delta_{\vec{k},\vec{k}'}\sum\limits_{F,\vec{k}''} \left( \delta_{A,C} V^{\vec{k},\vec{k}''}_{D,F}V^{\vec{k}'',\vec{k}}_{F,B} \delta(E_F^{k''}-E_D^{k} \pm \epq) \right. \\
&\left. \left. + \delta_{B,D} V^{\vec{k},\vec{k}''}_{A,F}V^{\vec{k}'',\vec{k}}_{F,C} \delta(E_F^{k''}-E_C^{k} \pm \epq) \right) \right] .
\end{split}
\end{equation}
Comparing the interaction picture Eq. \ref{eq:gen_so_interact} with our result \ref{eq:gen_so}, we see that the only difference is a permutation in the subband indices of the energy conserving delta functions (set in our derivation by the denominator in Eq. \ref{eq:dm_coh1}). This difference comes about because we are essentially using the Schrodinger picture in our derivation where all operators are time-independent. It can be verified that both equations give the same result for PP and PC rates, but differ for CP and CC couplings. It is interesting to note that a similar ambiguity arises if one attempts to reduce the NEGF equations to effective DM equations of motion.\cite{wacker_coherence_2008}

\begin{figure}[htb]
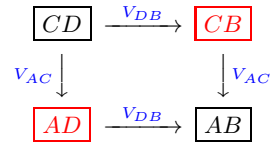

	\begin{equation*}
	\begin{CD}
	\boxed{CD} @>{\textcolor{blue}{V_{DB}}}>> \textcolor{red}{\boxed{CB}} \\
	@V{\textcolor{blue}{V_{AC}}}VV @VV{\textcolor{blue}{V_{AC}}}V \\
	\textcolor{red}{\boxed{AD}} @>\textcolor{blue}{V_{DB}}>> \boxed{AB}
	\end{CD}
	\end{equation*}
	\caption{Coherence-coherence scattering pathway from initial DM element (CD) to final DM element (AB) via scattering matrix elements $V_{AC}$ and $V_{DB}$ and intermediate coherences $CB$ and $AD$. The energy-conserving delta functions are associated with the indices of the (red) intermediate coherences in the Schrodinger picture and with the indices of the (blue) scattering processes in the interaction picture.}\label{fig:scatteringroutes}
\end{figure}
We can interpret this difference by loosely picturing each second-order superoperator element $\Gamma_{AB,CD}$ as a transition into an intermediate coherence followed by another scattering transition into the final element, as shown in Fig. \ref{fig:scatteringroutes}. (We examine a CC coupling as an example and suppress the $\vec{k}$ dependence for clarity.)
The energy conserving delta function in the interaction picture gets associated with the initial and final states of each ``leg'' of the scattering process ($AC$ or $DB$), whereas in the Schrodinger derivation it is set to the energy difference of the intermediate coherence between scattering events ($AD$ or $CD$). Physically the interaction picture approach suggests that the scattering events described by $V$ are relatively ``slow'' compared to the timescale of the Rabi oscillation between the scattering endpoints $\hbar/(E_D - E_B)$ or $\hbar/(E_A - E_C)$, whereas the Schrodinger picture implies that scattering is slow compared to the time scale of the intermediate coherence 
$\hbar/(E_A - E_D)$ or $\hbar/(E_C - E_B)$. We note that the Schrodinger picture approach has been previously studied in the literature for transport in coupled quantum dot systems, where it is sometimes referred to as the first-order von Neumann approach.\cite{goldozian_transport_2016} Such studies have also compared the effects of using the interaction and Schrodinger pictures and found any quantitative differences to be small.

\section{Superoperator Form for Specific Scattering Mechanisms}\label{sec:scattering}
Having established the basic structure of the generalized superoperator, we can proceed to calculate rates for physical scattering mechanisms. We take advantage of the fact that most quantum well-based QCLs are translationally invariant along the transverse directions and have isotropic energy dispersions in $\vec{k}$ depending only on the magnitude $k$ and transverse energy $E_k$. 
If we make the effective mass approximation (EMA), the energy of each subband state $E_A^k = E_A + E_k$ where $E_A$ is the band edge energy of the subband and $E_k = \dfrac{\hbar^2 k^2}{2m^*}$ with effective mass $m^*$. Note that the EMA is valid at low energies and thus usually suitable for THz QCLs, but in MIR devices the higher energy states mean that nonparabolic corrections  may be important, which can then be incorporated using an energy-dependent mass or a multiband \textbf{k$\cdot$p} model.\cite{burnett_origins_2016} Though we will use the EMA throughout this paper, the general procedure presented below can be modified to account for nonparabolicity if necessary.

Because of the in-plane translational invariance, we also assume that the electronic distribution is isotropic and varies only with the magnitude $k$ (and thus $E_k$) of the transverse momentum and not direction, i.e., we need only solve for $\rho^{E_k}_{AB}$ for different values of $E_k$, rather than for each $\vec{k}$. The most common scattering mechanisms for QCLs are either elastic (such as ionized impurities, alloy disorder, and interface roughness) or exchange a fixed quantum of energy (LO polar phonons assuming a constant phonon dispersion, considering absorption and emission separately). As a result, for any $\vec{k}$, each term in $\Gamma^{\vec{k},\vec{k}'}$ will be nonzero only for the set of $\vec{k}'$ at a single value of $E_{k'}$, as can be seen from the delta functions in Eqs. \ref{eq:gen_so}-\ref{eq:gen_so_elastic}.

These approximations allow for an enormous simplification of the superoperator terms, which in many cases can be partially or fully evaluated analytically as a function of subband and transverse energy. In Section \ref{subsec:enres}, we outline results for these energy resolved superoperator terms. 
Still more simplification is possible if we further assume each subband is in local equilibrium; this thermally averaged case is discussed below in Section \ref{subsec:thermal}.

\subsection{Energy-Resolved Scattering Theory}\label{subsec:enres}
Due to the energy conservation considerations noted above, we need only consider the scattering superoperator summed over $\vec{k}'$, i.e.,
\begin{equation}\label{eq:gen_so_ksum}
\Gamma_{AB,CD}^{E_k} = \sum\limits_{\vec{k}'}\Gamma_{AB,CD}^{\vec{k},\vec{k}'} ,
\end{equation}
which is useful because the summation over $\vec{k}'$ can be expressed in terms of integrals over the magnitude $k'$ and angle $\theta$, the latter of which can often be performed analytically. 
We can rewrite the energy-resolved scattering superoperator using Eq. \ref{eq:gen_so} as
\begin{equation}\label{eq:gamma_compact_eres}
\begin{split}
\Gamma_{AB,CD}^{E_k} =& \V_{AC,DB}^{E_k}(AD) + \V_{AC,DB}^{E_k}(BC) \\
&- \sum_F (\delta_{BD} \V_{AF,FC}^{E_k}(FB) + \delta_{AC} \V_{DF,FB}^{E_k}(FA))
\end{split}
\end{equation}
where for inelastic scattering (the phonon emission and absorption cases are distinguished by the $\pm$ sign)
\begin{equation}\label{eq:V_inel}
\begin{split}
&\V_{JK,LM}^{E_k}(\pm XY) = \frac{\pi}{\hbar}\sum_{\vec{q},\vec{k}'} \left(n_{\vec{q}} + \frac{1}{2} \pm \frac{1}{2}\right) \\
&\times 
V^{\vec{k},\vec{k}'}_{J,K}V^{\vec{k}',\vec{k}}_{L,M} \delta \left(\Delta_{XY}^{\pm \vec{q}} + E_k - E_{k'} \right)
\end{split}
\end{equation}
and $\Delta_{XY}^{\pm \vec{q}} = E_X - E_Y \pm \epq$ denotes the difference of the subband band edge energies. The extension to elastic scattering is obvious by removing the summation over $\vec{q}$ and setting $\epq=0$. Evaluating $\V$ for arbitrary arguments therefore suffices to describe the superoperator. For common single particle scattering mechanisms the multiple integrations implied by the $\vec{k}'$ and $\vec{q}$ dependence of $\mathcal{V}$ can be simplified, as discussed below.

Notice the appearance of the energy-conserving delta functions within $\V$, which specify the values of $\vec{k}'$ involved in the scattering. When applying the superoperator to the density matrix, each $\V$ term in $\Gamma^{E_k}_{AB,CD}$ selects the value of $\rho_{CD}^{E_{k'}}$ to which it couples. Therefore, the DM evolution equation can be conveniently written as
\begin{equation}\label{eq:dm_energyresolved}
\dot{\rho}_{AB}^{E_k} = \frac{i}{\hbar}[E_B^k - E_A^k]\rho_{AB}^{E_k} + \sum\limits_{C,D} \Gamma_{AB,CD}^{E_k} \rho_{CD}^E
\end{equation}
where the superoperator term on the right hand side represents
\begin{equation}
\begin{split}
\Gamma_{AB,CD}^{E_k} \rho_{CD}^E =& \sum\limits_{\pm} \left[ \V_{AC,DB}^{E_k}(\pm AD) \rho_{CD}^{E_k+\Delta_{AD}^{\pm \vec{q}}} \right. \\
&+ \V_{AC,DB}^{E_k}(\pm BC) \rho_{CD}^{E_k+\Delta_{BC}^{\pm \vec{q}}} \\
-& \sum_F (\delta_{BD} \V_{AF,FC}^{E_k}(\pm FB) \rho_{CD}^{E_k+\Delta_{FB}^{\pm \vec{q}}} \\
&+\left. \delta_{AC} \V_{DF,FB}^{E_k}(\pm FA) \rho_{CD}^{E_k+\Delta_{FA}^{\pm \vec{q}}} ) \right]
\end{split}
\end{equation}
in the case of inelastic scattering. Notice that the transverse energy term of each DM element is selected by the delta function argument of Eq. \ref{eq:V_inel}. When implementing these equations in numerical form over a finite set of $\vec{k}'$ (and hence $E_{k'}$), the delta function may need to be discretized, which can be done in a variety of ways; the method we used for the numerical calculations in this paper is discussed elsewhere.\cite{burnett_design_2016}

\subsubsection{Ionized Impurities}
Doping profiles in QCLs generally vary along the growth direction $z$ with some dopant concentration $N(z)$ (in units of cm$^{-3}$). Assuming that the impurity distribution along the in-plane directions $x, y$ is uncorrelated, disorder averaging\cite{kohn_quantum_1957} of the scattering matrix elements leads to
\begin{equation}\label{eq:imp_baseline}
\sum\limits_{\vec{k}'}V_{J,K}^{\vec{k},\vec{k}'}V_{L,M}^{\vec{k}',\vec{k}} = \frac{\pi}{\hbar\mathcal{A}}\sum\limits_{\vec{k}'} \int dz N(z) \mathbb{V}^{\vec{k},\vec{k}'}_{J,K}(z) \mathbb{V}^{\vec{k}',\vec{k}}_{L,M}(z)
\end{equation}
where $\mathcal{A}$ is the in-plane area and $\mathbb{V}^{\vec{k},\vec{k}'}_{J,K}(z)$ is the 2-D Fourier transform of the scattering potential. (Hereafter, during evaluation of this and other such quantities we will use cylindrical coordinates to transform $\sum\limits_{\vec{k}'} \rightarrow \dfrac{\mathcal{A}}{4\pi^2} \int k' dk' d\theta$.) While the true impurity potential may be quite complex due to the inhomogeneous electronic screening, a screened Coulomb potential with inverse Debye screening length $\eta$ is assumed for simplicity\cite{nelander_temperature_2009}. We can express $\mathbb{V}$ in terms of the subband envelope functions $\chi_A(z)$, but it turns out to be more efficient to use the Fourier transform of the product of subband states
\begin{equation}
\Phi_{AB}(q_z) = \int e^{iq_z z} \chi_A^*(z) \chi_B(z) dz.
\end{equation}
In this way we can write the scattering potential
\begin{equation}
\mathbb{V}^{\vec{k},\vec{k}'}_{A,B}(z) = \frac{e^2}{2\pi\epsilon_s} \int dq_z \frac{\Phi_{AB}(q_z) e^{-iq_z z}}{\eta^2 + q_z^2 + |\vec{k}-\vec{k}'|^2}
\end{equation}
where $\epsilon_s$ is the static dielectric constant of the device material. Substituting into Eq. \ref{eq:imp_baseline}, using $|\vec{k}-\vec{k}'|^2 = k^2+k'^2 - 2kk' \cos \theta$, and integrating over $\theta$, we obtain
\begin{widetext}
\begin{equation}\label{eq:impV_eres}
\begin{split}
\V_{JK,LM}^{E_k}(XY) = &\frac{e^4\hbar}{32 \pi^2 m^* \epsilon_s^2} \int dE_{k'} \int N(z) dz \int dq_{z1} \int dq_{z2} \Phi_{JK}(q_{z1})\Phi_{LM}(-q_{z2}) e^{i(q_{z2} - q_{z1})z} \\
&\times G(E_k,E_{k'},E_{qz1},E_{qz2})  \delta(\Delta_{XY} + E_k - E_{k'}) ,
\end{split}
\end{equation}
where
\begin{equation}
\begin{split}
G(E_k,E_{k'},E_{qz1},E_{qz2}) = \frac{1}{E_{qz1} - E_{qz2}} \left( \frac{1}{\sqrt{(E_{\eta} + E_k + E_{k'} + E_{qz2})^2 - 4E_kE_{k'}} } 
- \frac{1}{\sqrt{(E_{\eta} + E_k + E_{k'} + E_{qz1})^2 - 4E_kE_{k'}} }\right) 
\end{split}
\end{equation}
\end{widetext}
and we define $E_{\eta} = \dfrac{\hbar^2 \eta^2}{2 m^*}$ and $E_{qz1,qz2} = \dfrac{\hbar^2 q_{z1,z2}^2}{2m^*}$.
When $q_{z1} = q_{z2} = q_z$, $G$ reduces to
\begin{equation}
G(E_k,E_k',E_{qz}) = \frac{E_{\eta} + E_{k} + E_{k'} + E_{qz}}{\left[(E_{\eta} + E_k + E_{k'} + E_{qz})^2 - 4E_k E_{k'}\right]^{3/2}}
.
\end{equation}
The double integral over $q_{z1}$ and $q_{z2}$ appears because of the spatial inhomogeneity of the doping profile. In the case that $N(z) = N_d$ is constant, we can see that the phase factor in Eq. \ref{eq:impV_eres} vanishes unless $q_{z1}=q_{z2}$, removing the $z$ and $q_{z2}$ integrals; however, this approximation may be quite inaccurate in QCLs where doping is generally highly localized to reduce dephasing, in which case the full expression Eq. \ref{eq:impV_eres} should be evaluated. In this equation and similar ones below, we retain the integral over $E_{k'}$ to emphasize its role and for clarity in numerical implementations when discretizing energy. When handled analytically, the delta function removes this integral and substitutes $E_k + \Delta_{XY}$ for $E_{k'}$ everywhere.

\subsubsection{Alloy Disorder}
A basic model for a single alloy scatterer is $V(\vec{r}) = \Xi a^{3} \delta(\vec{r})$, where $\Xi$ is the strength of the potential and $a$ is the lattice spacing. The effective concentration of the alloy is $n = \frac{1}{a^3}x(1-x)$ where $x$ is the alloy fraction. Because of the assumed locality of the potential, the matrix element is independent of momentum and the scattering rate is simply
\begin{equation}
\begin{split}
\V_{JK,LM}^{E_k}(XY) = &\frac{\Xi^2 a^3 x(1-x) m^*}{2 \hbar^3} \int dE_{k'} \delta(\Delta_{XY} + E_k - E_{k'})\\
&\times \int\limits_{z_1}^{z_2} dz \chi_J^*(z) \chi_K(z) \chi_L^*(z) \chi_M(z)
\end{split}
\end{equation}
where $[z_1, z_2]$ define the spatial limits of the alloy material region in the growth direction $z$.

\subsubsection{Interface Roughness}
The usual phenomenological model for interface roughness assumes that the scattering is proportional to a correlation function $f(q)$, typically taken to be Gaussian or exponential, dependent on the momentum transfer $q$ between the initial and final states. We adapt this model and assume that scattering at an interface located at $z_i$ with band offset $\Delta E_i$ is given by
\begin{equation}
\begin{split}
V^{\vec{k},\vec{k}'}_{J,K} V^{\vec{k}',\vec{k}}_{L,M} = \frac{f(|\vec{k}-\vec{k}'|)\Delta E_i^2}{\mathcal{A}} \chi_J^{*}(z_i) \chi_K(z_i) \chi_{L}^{*}(z_i) \chi_{M}(z_i)
.
\end{split}
\end{equation}
Let us assume a Gaussian correlation function $f(q) = \pi \Omega^2 \Lambda^2 \exp\left( -\dfrac{\Lambda^2 q^2}{4} \right)$, where $\Lambda$ is the correlation length and $\Omega$ is the average interface displacement. Then defining $G_{JKLM} = \Delta E_i^2 \chi_J^{*}(z_i) \chi_K(z_i) \chi_{L}^{*}(z_i) \chi_{M}(z_i)$, we find
\begin{equation}
\begin{split}
&\V_{JK,LM}^{E_k} = \frac{\pi m^* \Omega^2 \Lambda^2}{2\hbar^3} \int dE_{k'} \exp\left( -\frac{m^* \Lambda^2}{2\hbar^2} (E_k+E_{k'})\right) \\
&\times I_0\left( \frac{m^* \Lambda^2 \sqrt{E_kE_{k'}}}{\hbar^2} \right)  G_{JKLM} \delta(\Delta_{XY}+E_k - E_{k'})
\end{split}
\end{equation}
where $I_0(z)$ is the modified Bessel function of the first kind. Scattering between different interfaces is usually assumed to be uncorrelated, so the total rate is obtained by summing the contributions of each interface.

\subsubsection{Polar Optical Phonons}
Most QCLs at present are made using III-V semiconductors in which the dominant inelastic scattering mechanism is the Fr{\"o}hlich interaction from longitudinal optical (LO) phonons. This is a long range Coulomb interaction which may be screened by free electrons in the semiconductor. We again assume Debye screening with screening vector $\eta$ to obtain the scattering rate for emission/absorption
\begin{equation}
V_{JK}^{kk'}V^{k'k}_{LM} = \alpha^2\sum_{\vec{q}} \frac{q^2}{(q^2 + \eta^2)^2}
\Phi_{JK}(\pm q_z) \Phi_{LM}(\mp q_z)
\end{equation}
where $\alpha^2 = \dfrac{E_{LO}e^2(\epsilon_{\infty}^{-1} - \epsilon_{s}^{-1})}{4\pi \mathcal{A}}$ with LO phonon energy $E_{LO}$ and high frequency material permittivity $\epsilon_{\infty}$. Converting the summation over $\vec{q}=(q_x,q_y,q_z)$ into an integral over $q_z$, noting that momentum conservation leads to $|\vec{k}-\vec{k'}|^2 = q_x^2 + q_y^2$, and integrating over the scattering angle, we find that
\begin{equation}
\begin{split}
&\V_{JK,LM}^{E_k}(\pm XY) = \left(n_{LO} + \frac{1}{2} \pm \frac{1}{2} \right)\dfrac{E_{LO}e^2(\epsilon_{\infty}^{-1} - \epsilon_{0}^{-1})}{16 \pi \hbar} \\
&\int dE_{k'} \int dq_z \left( \frac{\beta(\beta+ E_{\eta}) - 4E_k E_{k'}}{((\beta + E_{\eta})^2 - 4E_k E_{k'})^{3/2}} \right. \\
&\left. \times \Phi_{JK}(\pm q_z) \Phi_{LM}(\mp q_z)\delta(\Delta^{\pm}_{XY}+E_k - E_{k'}) \right)
\end{split}
\end{equation}
for $\beta = E_k + E_{k'} + \dfrac{\hbar^2q_z^2}{2m^*}$, $E_{\eta} = \dfrac{\hbar^2 \eta^2}{2 m^*}$, and $n_{LO}$ equal to the equilibrium Bose-Einstein phonon occupation. 

\subsection{Thermally Averaged Scattering Model}\label{subsec:thermal}
The equations above allow computation of the scattering rates and hence the density matrix resolved in transverse energy, which can give significant insight into the internal details of the device. No assumptions about the electron distribution as a function of $E_k$ are made, and indeed we will see later that in practical devices such distributions may be highly nonequilibrium. 
However, we can still obtain new insights (and reduction of computational effort) if we assume that each DM element obeys a Boltzmann distribution so that $\rho_{AB}^{E_k} = \rho_{AB} \mathcal{F}_{AB}\exp(-E_k/k_B T_{AB})$, where $\mathcal{F}_{AB}$ is a scaling factor and $T_{AB}$ is the effective electron temperature of the corresponding population or coherence $AB$.

Let us normalize the distribution so that $\sum\limits_{\vec{k}} \rho_{AB}^{E_k} = \rho_{AB}$, in which case
\begin{equation}
\mathcal{F}_{AB} = \dfrac{2\pi \hbar^2}{m^* \mathcal{A} k_B T_{AB}} .
\end{equation}
If we return to Eq. \ref{eq:dm_resolved} and sum over $\vec{k}$, we obtain the evolution of the thermal averaged DM
\begin{equation}\label{eq:dm_averaged}
\dot{\rho}_{AB} = \frac{i}{\hbar}[H_B - H_A]\rho_{AB} +\sum\limits_{CD}\bar{\Gamma}_{AB,CD}\rho_{CD} .
\end{equation}
We therefore define the thermal averaged superoperator
\begin{equation}\label{eq:gamma_thermalav}
\begin{split}
\bar{\Gamma}_{AB,CD} =& \bV_{AC,DB}^{CD}(AD) + \bV_{AC,DB}^{CD}(BC) \\
&- \sum_F (\delta_{BD} \bV_{AF,FC}^{CD}(FB) + \delta_{AC} \bV_{DF,FB}^{CD}(FA))
\end{split}
\end{equation}
where
\begin{equation}\label{eq:scattV_so_average}
\begin{split}
\bV_{JK,LM}^{CD}(XY) = &\frac{\pi}{\hbar}\mathcal{F}_{CD}\sum_{\vec{k},\vec{k}',\vec{q},\pm} \left(n_{\vec{q}} + \frac{1}{2} \pm \frac{1}{2}\right) V^{\vec{k},\vec{k}'}_{J,K}V^{\vec{k}',\vec{k}}_{L,M} \\
&\times e^{-E_{k'}/k_B T_{CD}}\delta\left(\Delta_{XY}^{\pm \vec{q}} + E_k - E_{k'} \right)
\end{split}
\end{equation}
for inelastic scattering and the analogue for elastic processes immediately follows. The thermal averaged elements can therefore be found from the energy-resolved rates by integrating the Boltzmann distribution over $E_k$ and $E_{k'}$. We summarize the expressions for Eq. \ref{eq:scattV_so_average} for impurities, alloy disorder, interface roughness, and LO phonons in Appendix \ref{app:thermal_av}. In contrast to the energy resolved model (Eq. \ref{eq:dm_energyresolved}), there is no energy dependence of $\rho_{CD}$ which needs to be considered when considering scattering in Eq. \ref{eq:dm_averaged}, since all momentum/energy summations are completely done within $\bar{\Gamma}$.

While the thermal averaged model is convenient and compact, the assumption of a thermalized distribution for each subband and coherence may not be true in general. It is interesting that experimental analysis of photoluminescence data suggests thermalized hot electron distributions do exist in some QCLs.\cite{vitiello_measurement_2005}
Nonetheless, the choice of electron temperature for each DM element is basically phenomenological. We note that we have not yet considered inelastic electron-electron (e-e) scattering in our calculations;\cite{li_effects_2004,winge_simple_2016} while this mechanism can be included in principle, it significantly complicates computation in practice and is often neglected in DM models. Prior studies suggest that e-e scattering contributes to intrasubband thermalization, while it plays a more subtle role in coherence dephasing due to the preservation of subband coherence during scattering.\cite{li_effects_2004}
In general, whatever its exact quantitative magnitude, e-e scattering should contribute towards thermalization, localization, and dephasing, and thus it should tend to smooth out the I-V characteristics. Comparison of the thermal averaged and energy-resolved models can also give indications of how results differ if this is the case.

\section{Periodicity, Optical Field, and Velocity}\label{sec:period}
Thus far we have discussed the effects of scattering in terms of an arbitrary set of subband eigenstates. In practice, QCLs are constructed from repeated modules of quantum wells, so that the eigenstates of one module will be replicated periodically across other modules with appropriate shifts in energy due to the applied potential. Suppose each module is of length $L$ with an externally applied potential drop $U$ across it. We can expect the density matrix to have the block matrix form
\begin{equation}\label{eq:dm_blockform}
\begin{split}
\rho &= \begin{bmatrix}
\ddots & & \vdots & & \reflectbox{$\ddots$} \\
& \rho_{-1,-1} & \rho_{-1,0} & \rho_{-1,1} & \\
\cdots & \rho_{0,-1} & \rho_{0,0} & \rho_{0,1} & \cdots \\
& \rho_{1,-1} & \rho_{1,0} & \rho_{1,1} & \\
\reflectbox{$\ddots$} & & \vdots & & \ddots
\end{bmatrix}
\end{split} 
\end{equation}
where the subscripts denote the module index with the understanding that the indices for subband, $k$, etc., are contained within the block matrices $\rho_{\mu,\nu}$. Because of periodicity, the two-script notation for the DM can be reduced to a single index, e.g., $\rho_{\mu,\nu} = \rho_{\nu - \mu}$, so that Eq. \ref{eq:dm_blockform} becomes
\begin{equation}\label{eq:dm_block_periodic}
\begin{split}
\rho = \begin{bmatrix}
\ddots & & \vdots & & \reflectbox{$\ddots$} \\
& \rho_{0} & \rho_{1} & \rho_{2} & \\
\cdots & \rho_{-1} & \rho_{0} & \rho_{1} & \cdots \\
& \rho_{-2} & \rho_{-1} & \rho_{0} & \\
\reflectbox{$\ddots$} & & \vdots & & \ddots
\end{bmatrix}
\end{split}.
\end{equation}
$\rho_0$ describes the DM of an individual module (i.e., intramodule populations and coherences), while $\rho_{\mu}$ describes intermodule coherences between states of a module and those of its $\mu^{th}$ neighbor. Equations for the superoperator elements can be generalized by extending the subband index $A$ to include module number $\overset{A}{\mu}$. By tracking the indices and using the periodic and nonperiodic properties of $\rho$ and $V$ (discussed in more detail in Appendix \ref{app:periodicity}), we can write
\begin{equation}
\dot{\rho}_{\begin{subarray}{c}AB\\ \mu \end{subarray}}^{E_k} = 
\sum\limits_{C,D}\Gamma_{\begin{subarray}{c}AB,CD\\ \mu,\nu \end{subarray}}^{E_k}
\rho_{\begin{subarray}{c}CD\\\nu \end{subarray}}^E 
\end{equation}
using the same convention for the superoperator energy dependence as in Eq. \ref{eq:dm_energyresolved}. By tracking the scattering elements across modules, we can generalize Eq. \ref{eq:gamma_compact_eres} to 
\begin{widetext}
\begin{equation}\label{eq:gamma_compact_eres_period}
\begin{split}
\Gamma_{\begin{subarray}{c}AB,CD\\\mu,\nu \end{subarray}}^{E_k} =& \sum\limits_{\sigma} \left[ \V_{\begin{subarray}{l}A,C;D,B\\ \nu, \sigma ; \nu+\sigma, \mu+\nu \end{subarray}}^{E_k}(AD - \sigma U) +
\V_{\begin{subarray}{l}A,C;D,B\\ \sigma, \mu ; \mu+\nu, \sigma+\mu \end{subarray}}^{E_k}(BC + \sigma U) \right.\\
& \left. - \sum_F (\delta_{BD} \V_{\begin{subarray}{l}A,F;F,C\\ \sigma+\nu, \mu+\nu; \mu+\nu, \mu+\sigma \end{subarray}}^{E_k}(FB - \sigma U)
 + \delta_{AC} \V_{\begin{subarray}{l}D,F;F,B\\ \nu, \sigma ; \sigma, \mu \end{subarray}}^{E_k}(FA + \sigma U) \right]
 ,
\end{split}
\end{equation}
where
\begin{equation}
\V_{\begin{subarray}{l}J,K;L,M\\ \alpha, \beta ; \gamma, \delta \end{subarray}}^{E_k}(\pm XY - \sigma U) = \frac{\pi}{\hbar}\sum_{\vec{k}',\vec{q}} \left( n_{\vec{q}}+\frac{1}{2}\pm\frac{1}{2} \right) V^{\vec{k},\vec{k}'}_{\begin{subarray}{l}J,K\\ \alpha, \beta\end{subarray}}V^{\vec{k}',\vec{k}}_{\begin{subarray}{l}L,M\\ \gamma, \delta \end{subarray}} \delta \left(\Delta_{XY}^{\pm \vec{q}} + E_k - E_{k'} - \sigma U \right)
.
\end{equation}
\end{widetext}
All equations for the subband averaged quantities $\bar{\Gamma}$ and $\bV$ generalize in the same way with a corresponding addition in indices. This method allows for arbitrarily long-range coupling between modules, though in general $\nu$, $\mu$, and $\sigma$ can usually be evaluated over at most a single neighbor $\pm 1$. 
Note that while the $\nu$ and $\mu$ indices denote the starting and ending intermodule coherence, the summation $\sigma$ tracks scattering between modules. In particular, even if intermodule coherences are small, scattering to $\sigma = \pm 1$ is important because it describes intermodule transfer of intramodule quantities, e.g., how charge in a certain module scatters into an neighboring module. This is necessary for current flow within a periodic system, as otherwise the charge evolves in a closed loop within a module.

At this stage we have developed a general scattering superoperator which accounts for coherent effects and incorporates periodicity. An optical field can be included straightforwardly and nonperturbatively by further generalizing the DM and coherent Hamiltonian as functions of frequency $\omega$\cite{burnett_origins_2016}
\begin{align}
\rho &= \rho(0) + \rho(+\omega)e^{i\omega t} + \rho(-\omega)e^{-i\omega t} \\
H &= H(0) + H(+\omega)e^{i\omega t} + H(-\omega)e^{-i\omega t}
\end{align}
where $H(0)$ is the steady state band structure Hamiltonian and $H(\pm \omega) = qFz$ is the optical dipole Hamiltonian with electric field $F$ at frequency $\omega$. We can define a general coherent superoperator $\mathcal{L}\rho = \frac{1}{i\hbar}[H,\rho]$ so that the DM evolves in time as
\begin{equation}
(\mathcal{L} + \Gamma) \rho = \dot{\rho}.
\end{equation}
The scattering superoperator acts independently on each frequency component of $\rho$. Writing out the frequency dependent behavior explicitly we obtain in vectorized form
\begin{equation}\label{eq:vectorized_optical}
\begin{bmatrix}
\mathcal{L}(0) + \Gamma & \mathcal{L}(-\omega) & 0 \\
\mathcal{L}(+\omega) & \mathcal{L}(0) + \Gamma & \mathcal{L}(-\omega) \\
0 & \mathcal{L}(+\omega) & \mathcal{L}(0) + \Gamma
\end{bmatrix}
\begin{pmatrix}
\rho(-\omega) \\ \rho(0) \\ \rho(+\omega)
\end{pmatrix} = \begin{pmatrix}
-i\omega \rho(-\omega) \\ 0 \\ +i\omega \rho(+\omega)
\end{pmatrix}
\end{equation}
Here we write $\rho$ as a vectorized list of the unknowns in the three representative submatrices and the superoperators $\mathcal{L}$ and $\Gamma$ in corresponding matrix form. The equation is straightforwardly generalized for multiple frequencies.\cite{burnett_origins_2016} Once the scattering superoperator has been computed, the calculation of the steady state (and optical response, if so desired) of the DM follows the usual steps.

DC current can be computed from the steady state DM by taking the expectation value of the velocity operator. The coherent velocity operator is defined as $v_{coh} = \dfrac{i}{\hbar}[H,z]$ where $z$ is the position operator. However, there may also be incoherent contributions to the current through the scattering superoperator, which can be inferred from the DM time evolution as discussed in Appendix \ref{app:periodicity}. The end result is the definition of an incoherent velocity operator with matrix elements
\begin{equation}
v^{E_k}_{\begin{subarray}{l}AB\\ \mu \end{subarray}} = \sum\limits_{\nu}\mathcal{T}^{E_k}_{\begin{subarray}{l}AB;CD\\ \mu, \nu\end{subarray}}z_{\begin{subarray}{l}CD\\ \nu \end{subarray}}
\end{equation}
where
\begin{widetext}
\begin{equation}\label{eq:incohj_so}
\begin{split}
\mathcal{T}_{\begin{subarray}{c}AB,CD\\\mu,\nu \end{subarray}}^{E_k} =& \sum\limits_{\sigma} \left[ \V_{\begin{subarray}{l}C,A;B,D\\ \sigma, \nu ; \mu+\nu, \nu+\sigma \end{subarray}}^{E_k}(AD - \sigma U) \left(1+ \frac{\delta_{\nu0}\delta_{CD}(\sigma - \nu)L}{z_{\begin{subarray}{c}CC\\0\end{subarray}}}\right) +
\V_{\begin{subarray}{l}C,A;B,D\\ \mu , \sigma ; \sigma+\mu, \mu+\nu \end{subarray}}^{E_k}(BC + \sigma U) \left(1+ \frac{\delta_{\nu0}\delta_{CD}(\mu - \sigma)L}{z_{\begin{subarray}{c}CC\\0\end{subarray}}}\right) \right.\\
& \left. - \sum_F (\delta_{BD} \V_{\begin{subarray}{l}C,F;F,A\\ \mu+\sigma, \mu+\nu; \mu+\nu, \sigma+\nu \end{subarray}}^{E_k}(FB - \sigma U) \left(1+ \frac{\delta_{\nu0}\delta_{CD}(\mu - \nu)L}{z_{\begin{subarray}{c}CC\\0\end{subarray}}}\right)
+ \delta_{AC} \V_{\begin{subarray}{l}B,F;F,D\\ \mu , \sigma; \sigma, \nu \end{subarray}}^{E_k}(FA + \sigma U) \right] .
\end{split}
\end{equation}
\end{widetext}
Close examination shows a simple connection between like terms of $\Gamma$ and $\mathcal{T}$: they share the same factors of $\V$, only with swapped indices $\overset{AB}{\mu} \leftrightarrow \overset{CD}{\nu}$. Therefore, they can be computed simultaneously. In the examples discussed below, the incoherent current contributions are generally small compared to the coherent current, although in some kinds of devices these contributions may be important.\cite{burnett_density_2014} (An example of the latter is the case of a superlattice biased such that neighboring states are separated by the LO phonon energy; hopping transport then occurs via phonon scattering down the ``ladder'' of localized eigenstates which manifests as incoherent current in our formalism.)
Finally, optical properties like gain can be similarly computed from the expectation values of the velocity operator; fortunately only the DC coherent velocity (operating on the frequency dependent $\rho(\pm \omega)$) is necessary, because the ac velocity operator $v(\omega) = i[H(\omega), z]/\hbar$ vanishes for $H \propto z$. The DC incoherent velocity does not couple to the optical field and hence does not directly contribute to optical response.

\section{Examples and Application of Formalism}\label{sec:examples}
\subsection{Model Implementation for Devices}
Although the approach discussed in this paper diverges in some conceptual ways from conventional DM QCL models, the implementation differs mainly in determination of the scattering superoperator elements, each of which can be evaluated with comparable effort as a FGR rate. For convenience, we summarize the computational steps here.
\begin{enumerate}
\item Calculate the QCL band structure from the device Hamiltonian $H$ (including bias drop per module $U$) and select the set of $N$ subband eigenstates comprising a module. Choose number of modules for which to consider intermodule coherences $N_c$ ($\mu$, $\nu$) and intermodule transfer ($\sigma$).
\item 
If energy-resolved information is required, choose a set of $N_E$ transverse energies to calculate for the density matrix (which has a total number of elements $N_{\rho}=(2N_c+1)N^2N_E$). If the thermally averaged equations are used, only the electron temperatures need to be specified for the $N_{\rho}=(2N_c+1)N^2$ DM elements.
\item
Compute the scattering superoperator $\Gamma$ between each DM element for the mechanisms of interest using Eq. \ref{eq:gamma_compact_eres_period} for energy-resolved calculations or Eq. \ref{eq:gamma_thermalav} for thermal averages, generalized for periodicity. The incoherent current superoperator $\mathcal{T}$ can be constructed simultaneously using Eq. \ref{eq:incohj_so}.
\item 
If optical properties are to be studied, include the optical field Hamiltonian in the time evolution and incorporate frequency-dependent elements of the DM as in Eq. \ref{eq:vectorized_optical}. Impose the condition that the sum of steady-state populations $\Tr(\rho(0))=1$ (this can replace any one of the equations for a steady-state population). Invert the system of equations to obtain the solution of $\rho$, from which charge and current densities, optical response, etc., can be extracted.
\end{enumerate}

Of course, it is also possible to study the time-dependent DM behavior once the superoperator is computed by solving the time evolution equation for $\dot{\rho}$ directly. Space charge effects can also be incorporated at the mean field (Hartree) level by solving the Poisson equation using the DM electron density and substituting the resulting potential into $H$, iterating until a self-consistent solution is obtained. Similarly, a self-consistent calculation of optical field in an operating laser device can be obtained by iterating the optical intensity for an assumed threshold gain.

\subsection{Example: Localization and Tunneling in Superlattices}
\begin{figure}[htb]
	\includegraphics[width=0.42\textwidth]{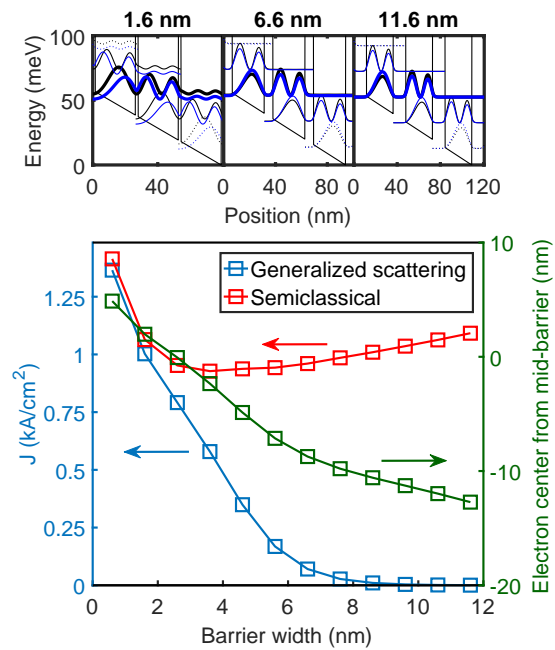}
	\caption{Top: GaAs/Al$_0.15$Ga$_0.85$As superlattice band structure as function of barrier thickness. Thick lines denote the two wave functions comprising each module; thin lines represent the shifted wave functions associated with neighboring modules. Bottom: current density (left axis) and degree of charge localization (right axis) versus barrier width. An uniform doping profile $N_d = 10^{16}$ cm$^{-3}$ is assumed and lattice temperature is 77 K.}
	\label{fig:twolevel}
\end{figure}

As a simple test of our model, and in order to demonstrate that it captures important coherence effects, we first consider a superlattice biased such that the ground state of each well is at resonance with the first excited state of the next. If we take into account only these two levels, this is a system of two subbands per period. (There are in fact other states in the superlattice, but they can be neglected for the sake of this demonstration.) We examine how this simple periodic two level system behaves as the barrier thickness between the wells is increased. The bandstructures we consider are shown in Fig. \ref{fig:twolevel}. The well length is held constant at 24.8 nm to give a separation of approximately 20 meV between the ground and first excited states in each well, while the barrier width is tuned from 0.6 to 11.6 nm. Even though the anticrossing condition becomes highly sensitive to field with thicker barrier, we can still always find a bias where the wavefunctions look approximately the same. The module energy drops are all close to 20 meV, but are adjusted slightly to account for the small Stark shifts that would otherwise move the states out of resonance.

In Fig. \ref{fig:twolevel}, we compare the current from our generalized scattering DM formalism with that obtained using the semiclassical approach (where only PP/FGR terms are used and all coherences are set to zero). We observe a striking disparity: whereas the full model predicts a strong decrease in the current with increasing barrier thickness, the semiclassical model predicts almost no change at all. The failure of the semiclassical model in this situation is well known and often cited as an example of why localized basis DM models are needed.\cite{callebaut_importance_2005} The cause of the failure is clear by inspection of the wave functions: because the energies and wave functions hardly change with the barrier, FGR rates using these basis states change very little as the barrier thickness is increased. Physically, this is because FGR transitions predict that charge scattering in from the previous module is instantaneously ``placed'' in an anticrossed eigenstate with equal density on either side of the barrier, with no delay due to tunneling. On the other hand, the generalized approach calculation allows scattering to induce coherence between the two anticrossed states, which allows for electron buildup behind the barrier, similar to the discussion of scattering terms in Section \ref{sec:relationship} and the illustration in Fig. \ref{fig:threelevelstates}. The coherent Hamiltonian $H$ then relaxes this coherence because of the state energy difference, amounting to tunneling. As the tunnel barrier thickens and this energy difference reduces, the strength of the tunneling weakens and current drops. This is equivalent to the behavior observed in the ``incoherent tunneling'' regime by Sirtori et al.\cite{sirtori_resonant_1998} Localization can also be quantified directly by taking the expectation value of the electron position within a single module using Tr($\rho_0 z_0$), as shown by the green curve and the right-side axis in Fig. \ref{fig:twolevel}. We see that the electron localizes increasingly to the negative (upstream) side of the barrier as the width is increased, indicating a significant buildup of charge with decreased tunneling.

\subsection{Application to Resonant Phonon THz QCL}
\begin{figure}[htb]
	\includegraphics[width=0.45\textwidth]{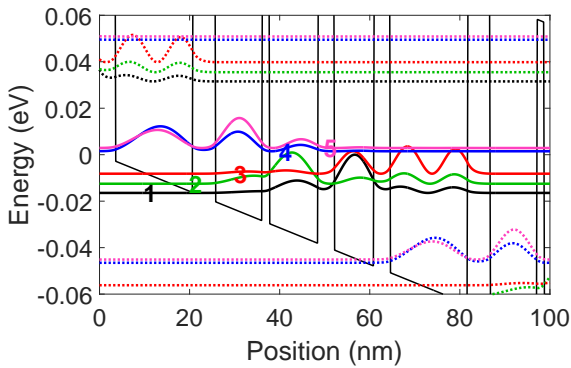}
	\caption{Band structure for diagonal resonant phonon THz QCL in GaAs/Al$_{0.15}$Ga$_{0.85}$As for module bias of 48 mV. The device layer thicknesses are 3.7/17.2/5.1/10.3/1.7/10.7/3.7/8.8 in nm. Solid lines indicate the eigenstates for a given module with labels indicated at right; dotted lines indicate the eigenstates for neighboring modules. States are labeled in ascending order of energy for convenience when describing different biases.}
	\label{fig:rpcdevice}
\end{figure}
\subsubsection{Device Simulation and Model Comparisons}
We turn our attention to a realistic QCL structure, a five-level diagonal resonant phonon design at $\sim3.4$ THz which previously produced a record 1.01 W in pulsed mode at 10 K.\cite{li_terahertz_2014} The eigenstates of the module near the design bias are illustrated in Fig. \ref{fig:rpcdevice}. The design uses a single injector well followed by a diagonal radiative transition (injector and upper states 4 and 5 in the figure). Extraction occurs through the strongly coupled three-state miniband (states 1-3), after which LO phonon emission occurs into the injector state of the next module. Localization and tunneling effects in this device make the semiclassical theory highly suspect, as we will see. At the same time, the complicated nature of the states makes it very difficult to choose a localized basis and construct a phenomenological DM theory for this device. (This does not necessarily mean that no such suitable set of states can be found, but that the choice is not obvious and, equally importantly, may not be robust against slight changes in the basis or in the design.) This design is therefore a good example of a practically relevant system not amenable to the usual modeling methods.

We therefore apply our theory to this design, using both the energy-resolved and thermal averaged methods. Conventional values for the interface roughness ($\Omega=2$\AA \space and $\Lambda=10$ nm) are used and a lattice temperature of 100 K is assumed. Since the scattering superoperator is calculated using known mechanisms, there are no other arbitrary fitting parameters such as dephasing times in the energy-resolved model. For the thermal averaged model, we assume that all DM elements share the same electron temperature $T_e$=100 K.

\begin{figure}[htb]
	\centering
		\includegraphics[width=0.45\textwidth]{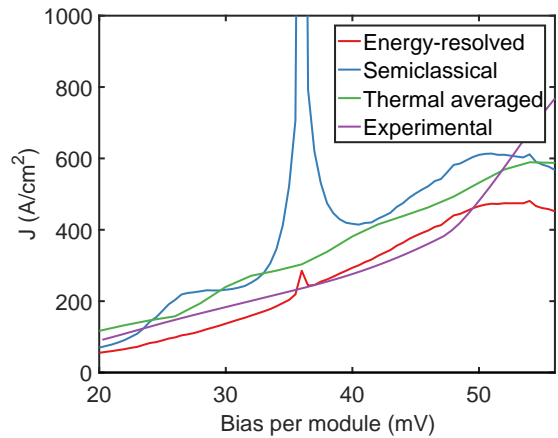}
		\caption{Current density versus module bias for experimental resonant phonon (dotted line) compared with semiclassical, energy-resolved, and thermal averaged DM models. The lattice temperature is assumed to be 100 K in all calculations, and the electron temperature is also 100 K in the thermal averaged model.}
		\label{fig:jcomp}
	
\end{figure}
In Fig. \ref{fig:jcomp} we compare the $J-V$ curves obtained using our calculation with the semiclassical prediction as well as experimental data measured in our lab from a device with this design. The module bias for the experimental curve is obtained by subtracting a 0.8 V Schottky bias drop from the experimentally applied voltage and dividing the resulting value by the number of periods (263); no additional series resistances or other parasitics are considered. The experimental design lases above 48 mV, which accounts for the increased current due to stimulated emission above threshold. The calculated currents shown are computed assuming zero optical excitation, so we focus on the subthreshold behavior here. Given the neglect of self-consistent electrostatics and explicit electron-electron scattering in the model, as well as the litany of experimental uncertainties in material quality, parasitic resistances, etc., the primary purpose of this comparison is to show that the model is physically consistent rather than to quantitatively fit experiment. In this regard, it is notable that the semiclassical theory inaccurately predicts strong negative differential resistance at several resonant peaks below threshold, peaks which are not observed in experiment. Importantly, we observe that these peaks are suppressed by localization effects once coherences are taken into account, as seen by their absence in the energy-resolved and thermal averaged calculations. The magnitude of the computed currents from the generalized scattering approach is also closer to experiment, keeping in mind the caveats about quantitative fits mentioned above. More detailed comparisons and discussion of experimental devices will be given elsewhere.\cite{burnett_notitle_nodate}

\begin{figure}[htb]
	\includegraphics[width=0.48\textwidth]{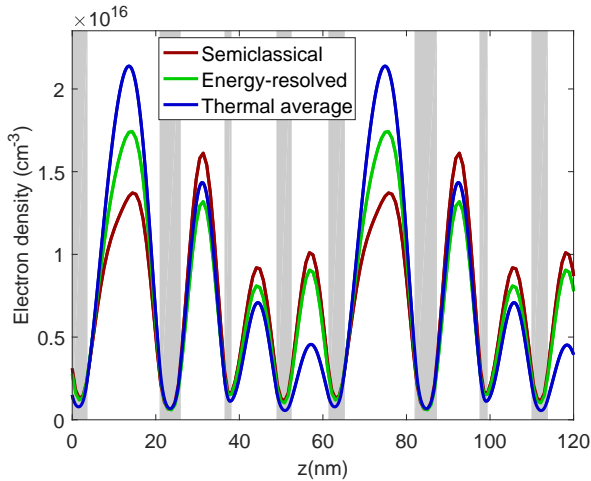}
	\caption{Electron density at 48 mV from semiclassical, energy-resolved, and thermal averaged calculations. The position axis corresponds to that in Fig. \ref{fig:rpcdevice}; barrier regions are shaded gray.}
	\label{fig:rpcdensity}
\end{figure}
\subsubsection{Interpretation of Coherences}
The effect of the coherences in the device can be seen directly by examining the calculated space charge densities in Fig. \ref{fig:rpcdensity}. We observe that the semiclassical calculation gives a higher charge density in the upper state well downstream from the injector, whereas both the energy-resolved and thermal averaged calculations lead to charge localization behind the barrier, as expected on physical grounds. This arises from the coherence between states 4 and 5, which can be seen from the thermally averaged density matrix in Table \ref{tab:rpcdm}.
\begin{table}
	\caption{\label{tab:rpcdm}Computed thermal averaged density matrix (magnitudes) for the resonant phonon structure at 48 mV assuming electron temperature of 100 K for all DM elements. The normalization is chosen so that Tr($\rho$)=1.}
	\begin{ruledtabular}
		\begin{tabular}{c | c c c c c}
			\bf{State} & \bf{1} & \bf{\textcolor{green}{2}} & \bf{\textcolor{red}{3}} & \bf{\textcolor{blue}{4}} & \bf{\textcolor{CarnationPink}{5}} \\ \hline
			\bf{1} & 0.0921	& 0.0085 & 0.00138 & 0.00089 & 0.0005 \\
			\bf{\textcolor{green}{2}} & 0.0085	& 0.0602	& 0.008 &	0.0018 &	0.0006 \\
			\bf{\textcolor{red}{3}} & 0.00138	& 0.008	& 0.0233 &	0.0053 &	0.001 \\
			\bf{\textcolor{blue}{4}} & 0.00089 & 0.0018	& 0.0053 &	0.486	& 0.107 \\
			\bf{\textcolor{CarnationPink}{5}} & 0.0005	& 0.0006	& 0.001 &	0.107 &	0.338
		\end{tabular}
	\end{ruledtabular}
\end{table}
Here the populations of states 4 and 5 are unsurprisingly the largest, showing the population inversion characteristic of lasers. However, the magnitude of the coherence between these states $\rho_{45}$ is also significant, giving rise to the strong localization of charge behind the relatively thick injection barrier. This is by far the largest coherence, though $\rho_{23}$ is also a sizeable fraction of the related state populations, showing the coupling between the extraction minibands. 

\begin{table}
	\caption{\label{tab:rpcso}Selected intramodule superoperator elements at 48 mV module bias assuming thermal averaged electron temperature of 100 K.}
	\begin{ruledtabular}
		\begin{tabular}{c  c | c c}
			$\bar{\Gamma}_{4,4}$ & $-2 \times 10^{12}$ s$^{-1}$ &
			$\bar{\Gamma}_{45,45}$ & -$4.3 \times 10^{12}$ s$^{-1}$ \\
			$\bar{\Gamma}_{5,5}$ & $-2.4 \times 10^{12}$ s$^{-1}$ &
			$\bar{\Gamma}_{45,5}$ & $1.01 \times 10^{12}$ s$^{-1}$ \\
			$\bar{\Gamma}_{4,5}$ & $2.26 \times 10^{12}$ s$^{-1}$ & 
			$\bar{\Gamma}_{45,4}$ & -$9.3 \times 10^{11}$ s$^{-1}$ \\
			$\bar{\Gamma}_{5,45}$ & $1.43 \times 10^{12}$ s$^{-1}$ &
			$\bar{\Gamma}_{45,54}$ & $2.77 \times 10^{12}$ s$^{-1}$ \\
			$\bar{\Gamma}_{4,3}$ & $1.22 \times 10^{12}$ s$^{-1}$ &
			$\bar{\Gamma}_{45,3}$ & -$6.9 \times 10^{11}$ s$^{-1}$ \\
			$\bar{\Gamma}_{23,2}$ & $9.9 \times 10^{11}$ s$^{-1}$ &
			$\bar{\Gamma}_{23,23}$ & -$3.6 \times 10^{12}$ s$^{-1}$
		\end{tabular}
	\end{ruledtabular}
\end{table}
To identify the origin of these coherences in our calculations, we list some of the most important thermally averaged superoperator terms in Table \ref{tab:rpcso}. The dephasing times $\bar{\Gamma}_{45,45}$ and $\bar{\Gamma}_{23,23}$ are largest in magnitude, though interestingly we can also observe large ``adjoint CC'' terms like $\bar{\Gamma}_{45,54}$; since the DM is Hermitian, these terms counteract dephasing to some degree. The buildup of $\rho_{45}$ and $\rho_{23}$ is largely driven by PC rates like $\bar{\Gamma}_{45,5}$ and $\bar{\Gamma}_{23,2}$; importantly, we see that these terms are of the same order of magnitude as FGR-type PP transitions such as $\bar{\Gamma}_{4,5}$. Note also the sign difference of $\bar{\Gamma}_{45,5}$ and $\bar{\Gamma}_{45,4}$, reflecting how each interaction acts to localize the density on either side of the injection barrier. (Recall from the discussion in Section \ref{sec:features} that the signs of such terms and their ``direction of localization'' must be interpreted considering the particular phases of the module wave functions.)

\begin{figure*}[htb]
	\centering
	\begin{subfigure}[t]{0.32\textwidth}
	\includegraphics[width=1.05\textwidth]{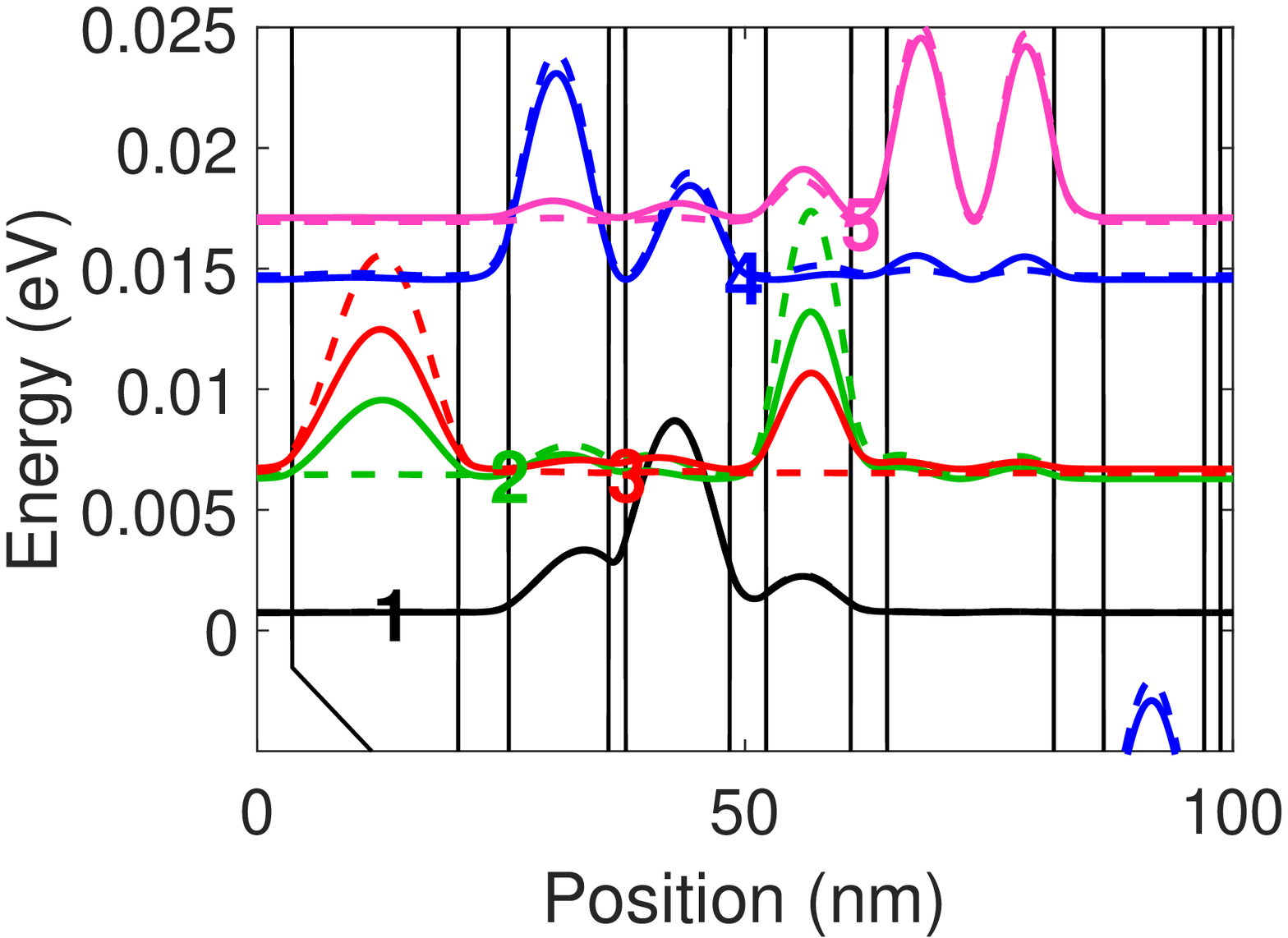}
	\caption{}
	\label{fig:dmeig26}
	\end{subfigure}
	\begin{subfigure}[t]{0.32\textwidth}
	\includegraphics[width=1.05\textwidth]{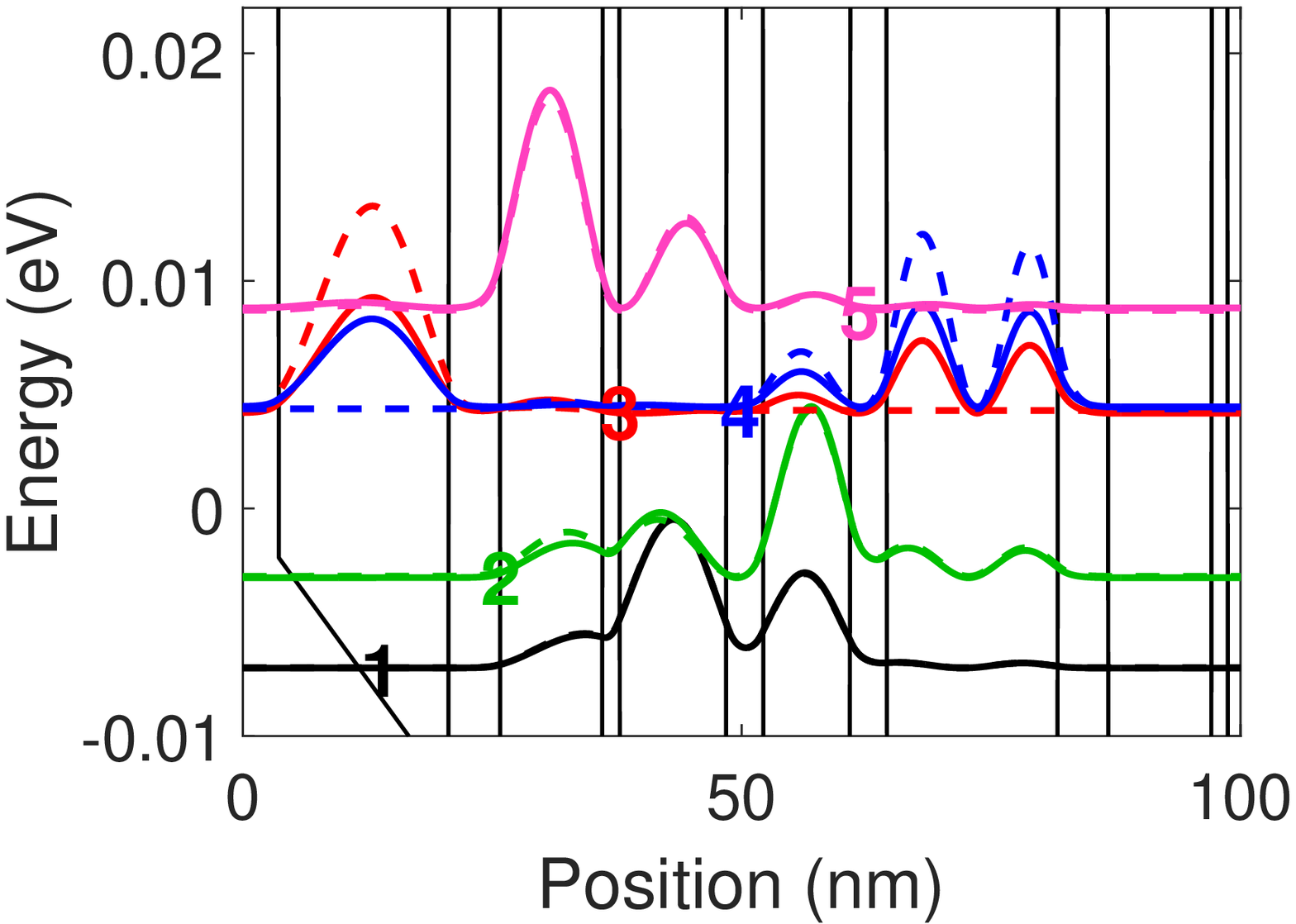}
	\caption{}
	\label{fig:dmeig36}
	\end{subfigure}
	\begin{subfigure}[t]{0.32\textwidth}
		\includegraphics[width=1.05\textwidth]{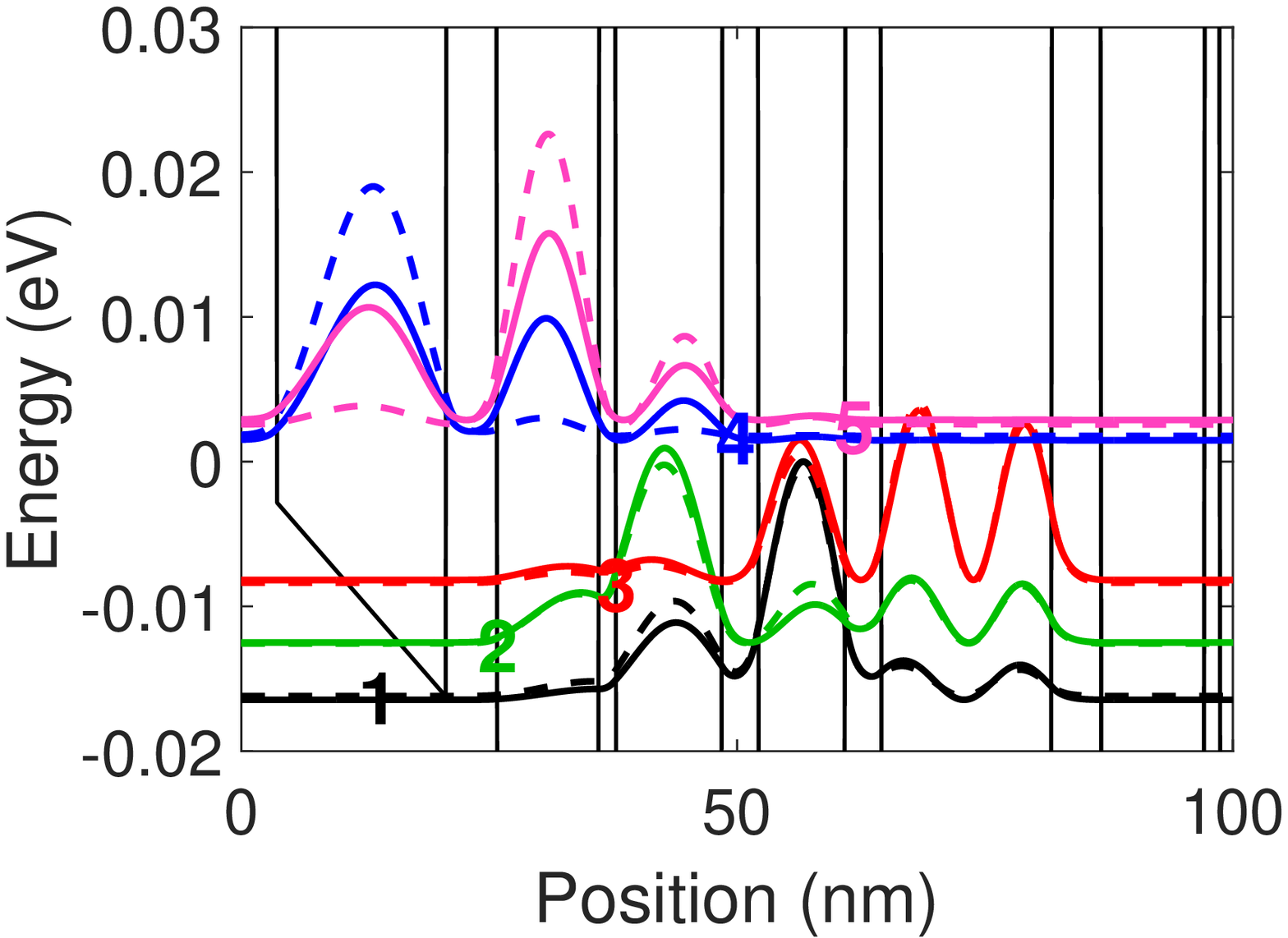}
		\caption{}
		\label{fig:dmeig48}
	\end{subfigure}
	\caption{Probability densities of energy eigenstates (solid lines) and density matrix eigenstates (dashed lines) at (a) 26 mV, (b) 36.7 mV, and (c) 48 mV module bias. Since the density matrix eigenstates do not have definite energies, they are plotted on the y-axis with respect to their energy expectation values.}
	\label{fig:dmeig}
\end{figure*}
\begin{table}
	\caption{\label{tab:dmeig}Populations $\rho$, coherent velocities $v_{c}$, and incoherent velocities $v_{ic}$ corresponding to the eigenstates of the thermal averaged density matrix at 26, 36.7, and 48 mV module bias (states correspond to those in Fig. \ref{fig:dmeig}). Populations are normalized to 1; velocities are in units of $10^4$ cm/s.}
	\begin{ruledtabular}
		\begin{tabular}{c | c c c c c}
			 & \bf{1} & \bf{\textcolor{green}{2}} & \bf{\textcolor{red}{3}} & \bf{\textcolor{blue}{4}} & \bf{\textcolor{CarnationPink}{5}} \\ \hline
			$\rho$(26) & 0.267 & 0.103 & 0.575 & 0.043 & 0.013 \\
			$v_{c}$(26) & -4.9 & 20	& 18 & 81 &	-110 \\
			$v_{ic}$(26)  & 0.37 & 4.8 & -0.46 & 0.89 & 15 \\
			\hline
			$\rho$(36) & 0.162 & 0.084 & 0.63 & 0.017 & 0.11 \\
			$v_{c}$(36) & 51 & 56 & 24 & -139 &	9.0 \\
			$v_{ic}$(36)  & 1.3	& 5.5 & -0.58 & 20 & 1.8 \\
			\hline
			$\rho$(48) & 0.094 & 0.06 & 0.022 & 0.54 & 0.28 \\
			$v_{c}$(48) & 269 & -5 & -289 & 61 & -36 \\
			$v_{ic}$(48)  & 5.0	& 7.9 & 2.0 & -0.7 & 1.8 \\
		\end{tabular}
	\end{ruledtabular}
\end{table}

In our discussion of the desired operating point (48 mV) for this device, the coherence between the injector and upper level is most important, as expected. However, in general different coherences may become important (and indeed the underlying states themselves may change) as a function of bias and device design. One way to gauge these effects is to compute the DM eigenstates. The corresponding eigenvalues are simply the occupation probabilities (populations) of these states. Because there are no off-diagonal elements of $\rho$ in this basis, the current ``carried'' by each state is given simply by the product of its population and the corresponding diagonal element of the velocity operator in this basis. In Fig. \ref{fig:dmeig}, we compare the energy eigenstates and DM eigenstates at three different biases (26 mV, 36.7 mV, and 48 mV), each corresponding to a different anticrossing in the structure (which gives rise to the resonant spikes in the semiclassical current). We use the thermally averaged density matrix for simplicity to avoid any possible complications from the $E_k$ dependence of the energy-resolved DM.

As the bias increases, successive anticrossings occur between the injector well and adjacent wells, as seen in the progression from Figs. \ref{fig:dmeig}a-c. In the semiclassical approach, scattering ``instantaneously'' transfers charge between these wells, leading to the sudden spikes in current around these bias points in Fig. \ref{fig:jcomp}. However, the localizing effects of PC and CC transitions in our generalized scattering approach are evident in how the delocalized stationary states split into separate states strongly localized in one well.

It is also of interest to examine the relative populations and associated velocities of the DM eigenstates, which are listed in Table \ref{tab:dmeig}. In each case, as expected, the most highly occupied state is the one localized in the injector well (state 3 at 26 and 36.7 mV and state 4 at 48 mV). The associated coherent ``velocity'' of this state, which can be physically interpreted as its tunneling rate, increases with bias because 1) the tunneling probability rises with field and 2) the effective tunneling distance reduces as the separation to its anticrossed pair state reduces. Note that the effective velocities of the other states differ in magnitude and can even become negative, where the latter indicates that the net tunneling out of the state occurs against the direction of the field (generally because there are no states downstream which are near tunneling resonance). An example of this is the negative velocity of DM eigenstate 5 at 48 mV (pink dashed line in Fig. \ref{fig:dmeig48}), which reflects the possibility of tunneling from the upper state back into the injector state. This can be understood clearly by considering Rabi oscillations in a two-level system, as discussed in Appendix \ref{app:dmeig}. While some DM eigenstates will inevitably have negative coherent velocities, as discussed in the appendix, the current carried by these states can be minimized by reducing their population.

The higher velocities of the lower states are important in allowing efficient extraction of charge, though the small populations of these states means that their contribution to the total current is smaller than that of the injector state. In Table \ref{tab:dmeig} we also list the corresponding incoherent velocities of the DM eigenstates; in general we can see that these are much smaller than the coherent velocities, reflecting the minor contribution scattering makes directly to the current. An exception occurs for state 2 at 48 mV module bias (an extractor state) where the positive incoherent velocity actually exceeds the coherent velocity, probably due to phonon depopulation of the extractor occurring faster than backwards resonant tunneling. In general, scattering plays a critical role in transport in our theory, but usually indirectly by inducing coherences and hence coherent current. We note again that this interpretation of the device DM emerges naturally without the need to separately define localized states or tunnel couplings for each bias; indeed the DM eigenstates better reflect the ``true'' localization of electrons induced by the kinetics of the system.

\begin{figure}[hbt]
	\includegraphics[width=0.4\textwidth]{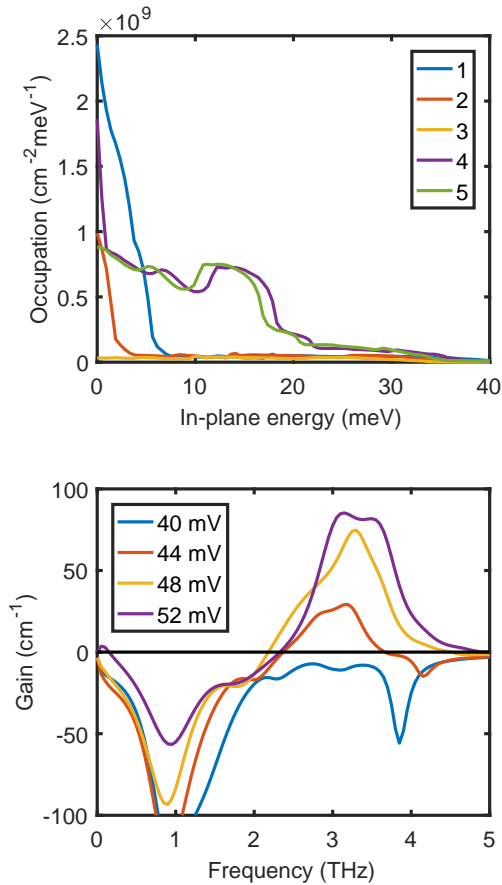}
	\caption{(a) Energy-resolved distributions within subbands at 48 mV (subband labels correspond to wave functions in Fig. \ref{fig:rpcdevice}). (b) Small signal gain as function of frequency and bias.}
	\label{fig:rpcdist}
\end{figure}

\subsubsection{In-Plane Energy Distributions and Optical Response}
Naturally, the quantitative values of the density matrix and the superoperator elements are strong functions of bias, temperature, etc., so that the most important rates and coherences may change at different operating points and depend on whether the distributions are thermalized. For example, in Fig. \ref{fig:rpcdist}a we plot the subband populations as function of in-plane energy calculated using the energy-resolved model. While the previously discussed caveats about the absence of e-e scattering hold, we observe that the injector and upper states 4 and 5 have strongly nonthermal distributions, in agreement with prior observations.\cite{jonasson_partially_2016} However, we also observe that the upper states are much ``hotter'' than the lower extraction states; in particular, the distribution for state 1 can be fitted quite well with a Boltzmann distribution with electron temperature of 20 K, much lower than the lattice temperature of 100 K. This occurs because electrons in subband 1 with transverse energies greater than about 10 meV can emit an optical phonon and scatter into states 4 or 5 in the next module; however, cold electrons at the bottom of the band cannot scatter out except via phonon absorption, a weak process at low lattice temperature. As a result, charge builds up at the bottom of subband 1; in fact, about 20\% of electrons are found in that band in the energy-resolved calculation compared to about 9\% in the thermally averaged DM in Table \ref{tab:rpcdm}. This accounts for a significant difference in the predictions of the energy-resolved and thermal averaged calculations for the shape of the charge density in the lower/extractor wells in Fig. \ref{fig:rpcdensity}; the increased buildup of charge in the fourth well in the energy-resolved case is due to the higher population of state 1, which is centered in that well as seen in Fig. \ref{fig:rpcdevice}.

Finally, as an illustration of optical response, in Fig. \ref{fig:rpcdist}b we plot the small signal gain as a function of frequency and bias predicted by the model. As expected we observe absorption at low and high frequencies with gain peaked around 3-3.5 THz, increasing with bias. We note that our calculations are not limited to small signal quantities; studies of the effect of strong field intensity and detailed discussion of the optical response of the device will be given elsewhere.
The results presented here show that our approach can give detailed and quantitatively meaningful descriptions of devices which cannot be described well using conventional rate equations or density matrix models. Because it does not require divination of ad hoc localized basis states for each device, this approach can be directly applied to different categories of design. Additional calculations and analysis for the resonant phonon device shown here, as well as applications of the model to a variety of other THz QCL designs, will be discussed in a separate study.\cite{burnett_notitle_nodate}

\section{Conclusion}
We have derived a generalized density matrix approach suitable for studying the steady-state and optical properties of QCLs. The model eliminates the need for choosing suitable localized wave functions for a particular device by using the well-defined energy eigenstates, while accounting for coherences through a generalized scattering superoperator. Both energy-resolved and thermally averaged versions of the theory are presented. We demonstrate that this model explains and reproduces the spatial localization and tunneling behavior important for describing QCLs without the need for adopting an arbitrary tight-binding basis. We study a resonant phonon THz QCL device using our approach, showing in particular how the additional scattering pathways suppress subthreshold current peaks in closer agreement with experiment. By examining the form of the density matrix, its eigenstates, and the associated velocities, we gain a stronger intuitive understanding of device operation.

\begin{acknowledgments}
We would like to thank Prof. A. Wacker for useful discussions concerning the Schrodinger and interaction representations and the first-order von Neumann approach. This work was partially supported by NSF Grants 1150071 and 1509801.
\end{acknowledgments}

\appendix
\section{Matrix Notation for Superoperator}\label{app:matnote}
The master equation for the density matrix Eq. \ref{eq:dm2ndorder1_electron} may be symbolically expressed in a compact notation using a matrix definition of the delta function $[\delta E]_{ab} = \delta(E_a - E_b)$ and the Hadamard operation $\circ$ which performs element-wise multiplication, i.e., each element of the Hadamard product of two matrices $[A \circ B]_{ab} = A_{ab}B_{ab}$. The final DM equation is summarized as
\begin{equation}\label{eq:DM_gen}
\begin{split}
\dot{\rho} = & \frac{i}{\hbar}[\rho, H] + \frac{\pi}{\hbar}\sum_m [\delta E \circ [V_m, \rho], V_m] \\
+ \frac{\pi}{\hbar} \sum_{q\pm} & \left(n_q + \frac{1}{2} \pm \frac{1}{2}\right)[\delta(E \pm E_q) \circ (V_q \rho) \\
&- \delta(E \mp E_q) \circ (\rho V_q), V_q]
\end{split}
\end{equation}
where $V_m$ is an elastic scattering mechanism and $V_q$ is an inelastic scattering mechanism with mode $q$. This expression is valid for a general electronic system. The notation also makes clear that the derived superoperator is not Lindbladian.\cite{lindblad_generators_1976}

It is also interesting to contrast this with the interaction representation result (discussed in more detail in Section \ref{subsec:schrodint}), which in our notation is given by
\begin{equation}
\begin{split}
\dot{\rho} = &\frac{i}{\hbar}[\rho, H] + \frac{\pi}{\hbar}\sum\limits_m[[\delta E \circ V_m,\rho],V_m] \\
+ \frac{\pi}{\hbar} \sum_{q\pm} & \left(n_q + \frac{1}{2} \pm \frac{1}{2}\right)[(\delta(E\pm E_q)\circ V_q)\rho \\
&- \rho(\delta(E\mp E_q)\circ V_q), V_q] .
\end{split}
\end{equation}

\begin{widetext}
\section{Thermal Averaged Scattering}\label{app:thermal_av}
In this appendix we summarize the thermally averaged scattering superoperator terms for coupling between arbitrary DM elements, i.e., the form of Eq. \ref{eq:scattV_so_average} for the single particle scattering mechanisms considered in Section \ref{sec:scattering}. In the integrations over energy for a given term $\bV_{JK,LM}$, a cutoff is necessary depending on whether the band edge of $JK$ is below or above that of $KM$. In the following equations,  $\lambda$ describes the correction due to this cutoff such that $\lambda = 0$ if $\Delta_{XY}>0$ and $\lambda=-\Delta_{XY}$ if $\Delta_{XY}<0$.

\subsubsection{Ionized Impurities}
We can summarize the scattering rate for impurity scattering as
\begin{equation}
\begin{split}
\bV_{JK,LM}^{CD}(XY) = \frac{\mathcal{F}_{CD}\mathcal{A}e^4}{128\pi^3\hbar\epsilon^2}\exp\left(-\frac{\Delta_{XY}}{k_B T_{CD}}\right)\int dz N(z) \\
\times \int dq_{z1} \int dq_{z2} \Phi_{JK}(q_{z1})\Phi_{LM}(-q_{z2}) e^{i(q_{z2} - q_{z1})z} G(q_{z1},q_{z2})
\end{split}
\end{equation}
where
	\begin{equation}
	G(q_{z1},q_{z2})=\frac{\sqrt{\pi k_B T_{CD}}}{E_{qz1}-E_{qz2}}\left[ \frac{\exp\left(\frac{\Gamma_2}{k_B T_{CD}}\right) \text{erfc}\left(\sqrt{\frac{\Gamma_2 + \lambda}{k_B T_{CD}}}\right)}{\sqrt{E_{\eta} + E_{qz2}}} - \frac{\exp\left(\frac{\Gamma_1}{k_B T_{CD}}\right) \text{erfc}\left(\sqrt{\frac{\Gamma_1 + \lambda}{k_B T_{CD}}}\right)}{\sqrt{E_{\eta} + E_{qz1}}} \right]
	\end{equation}
which for the special case $q_{z1} = q_{z2} = q$ reduces to
	\begin{equation}
	\begin{split}
	G(q,q) &= \frac{\exp(\Gamma/k_B T_{CD})}{2(E_{\eta} + E_q)^{3/2}} \times \\
	&\left[  \sqrt{\pi k_B T_{CD}}\left(1+\frac{2\Gamma - E_{\eta}-E_q - \Delta_{XY}}{k_B T_{CD}}\right)\text{erfc}\left( \sqrt{\frac{\Gamma+ \lambda}{k_B T_{CD}}}\right) + \frac{E_{\eta} + E_q + \Delta_{XY} - 2\Gamma}{\sqrt{\Gamma+\lambda}}\exp\left(-\frac{\Gamma+\lambda}{k_B T_{CD}} \right)\right]
	\end{split}
	\end{equation}
and we define for wave vector $q_n$
\begin{equation*}
	\Gamma_n = \frac{\left(E_{\eta} + \Delta_{XY} + E_{qn}\right)^2}{4(E_{\eta}+E_{qn})} .
\end{equation*}

\subsubsection{Alloy Scattering}
For alloy scattering we again assume a local scattering potential with matrix element $\Xi$ and lattice spacing $a$ within spatial region of alloyed material [$z_1$,$z_2$] with alloy fraction $x$. The averaged rate is then
\begin{equation}
\bV_{JK,LM}^{CD}(XY) = C_{all} \int\limits_{z_1}^{z_2} dz \chi_J^*(z) \chi_K(z) \chi_L^*(z) \chi_M(z)
\end{equation}
where
\begin{equation}
C_{all} = x(1-x)\frac{\Xi^2 a^3 m^{*2}\mathcal{A}\mathcal{F}_{CD}k_B T_{CD}}{4\pi\hbar^5}\exp\left(-\frac{\lambda+\Delta_{XY}}{k_B T_{CD}}\right) .
\end{equation}

\subsubsection{Interface Roughness}
As before, we take a Gaussian correlation function for the roughness profile and integrate to find
\begin{equation}
\bV_{JK,LM}^{CD}(XY) = \frac{m^{*2} \Omega^2 \Lambda^2 G_{JKLM} \mathcal{F}_{CD}\mathcal{A}}{4 \hbar^5} \int_{\lambda}^{\infty} \exp\left( -(E_k + \Delta_{XY})\left[\frac{1}{k_B T_{CD}} + \frac{\Lambda^2 m^*}{2\hbar^2} \right]\right) 
\times I_0\left( \frac{m^*\Lambda^2}{\hbar^2} \sqrt{ E_k(\Delta_{XY} + E_k) }\right) dE_k
\end{equation}
Here $G_{JKLM} = \Delta_i^2 \chi_J^{*}(z_i) \chi_K(z_i) \chi_{L}^{*}(z_i) \chi_{M}(z_i)$ at the $i^{th}$ interface with band offset $\Delta_i$. Assuming the roughness profile across interfaces are uncorrelated, the total scattering rate is given by the sum of the rates for each interface.

\subsubsection{Polar Optical Phonons}
Assuming screened longitudinal optical phonon scattering, we find
	\begin{equation}
	\begin{split}
	\bV_{JK,LM}^{CD}(XY) &= \sum\limits_{\pm} \left( n_{LO} + \frac{1}{2} \pm \frac{1}{2}\right)\frac{\mathcal{A}\mathcal{F}_{CD} E_{LO} e^2 m^*}{128\pi^2 \hbar^3}\left( \epsilon_{\infty}^{-1} - \epsilon_{DC}^{-1} \right)\exp\left( \frac{-\Delta_{XY}^{\pm}}{k_B T_{CD}} \right) \int dq_z \Phi_{JK}(\pm q_z) \Phi_{LM}(\mp q_z) G_{\pm}(q_z) \\
	G_{\pm}(q_z) &= \frac{\exp\left(\dfrac{\Gamma_{\pm}}{k_B T_{CD}}\right)}{(E_{qz} + E_{\eta})^{3/2}} \left[  \frac{(E_{qz} + \Delta_{XY}^{\pm})(E_{qz} + \Delta_{XY}^{\pm} + E_{\eta}) - 2\Gamma_{\pm}(2E_{qz} + E_{\eta})}{\sqrt{\Gamma_{\pm} + \lambda_{\pm}}}\exp\left(-\frac{\Gamma_{\pm} + \lambda_{\pm}}{k_B T_{CD}} \right) \right. + \\
	&\left. \left( \{2E_{qz} + E_{\eta}\}\left\{ \sqrt{\pi k_B T_{CD}} + 2\Gamma_{\pm} \sqrt{\frac{\pi}{k_B T_{CD}}} \right\} - \{E_{qz} + \Delta_{XY}^{\pm}\}\{E_{qz} + \Delta_{XY}^{\pm} + E_{\eta}\}\sqrt{\frac{\pi}{k_B T_{CD}}}\right)\text{erfc}\left(\sqrt{\frac{\Gamma_{\pm} + \lambda_{\pm}}{k_B T_{CD}}}\right) \right]
	\end{split}
	\end{equation}
where
\begin{equation}
\Gamma_{\pm} = \frac{(E_{qz} + E_{\eta} + \Delta_{XY}^{\pm})^2}{4(E_{qz} + E_{\eta})}
\end{equation}
and $\lambda_{\pm} = 0$ if $\Delta_{XY}^{\pm} > 0$ and $\lambda_{\pm} = -\Delta_{XY}^{\pm}$ otherwise.

\end{widetext}

\section{Periodicity for Scattering and Velocity}\label{app:periodicity}
For periodic structures, the DM $\rho$, scattering potentials $V$, and other operators can be written in the block matrix form of Eq. \ref{eq:dm_blockform}. The formalism shown here can reach arbitrarily far from the diagonal, but will be truncated based on the energy selectivity and coherences spanning not more than one module (to be shown later). Periodic boundary conditions will be invoked to reduce the two-script notation to a single script ($\rho_{\sigma,\nu} = \rho_{\nu - \sigma}$) as in Eq. \ref{eq:dm_block_periodic}, but this cannot be done for $V$: since correlation is intended to be dropped between positions along $z$ in the scattering potential, the elements in $V$ are treated as functions of $z$ which are inner producted (not scalars), and therefore knowledge of the specific modules involved in $VV$ products must be retained. A concrete example is that $V$ might represent scattering from a single rough interface: $V_{0,1}$ and $V_{−1,0}$ will be different from each other because the states involved are in different locations relative to the particular interface.

Using a general notation for the DM basis states, we find
\begin{equation}
\begin{split}
\dot{\rho}_{\begin{subarray}{c}ab\\ 0 \mu \end{subarray}} &= \dot{\rho}_{\begin{subarray}{c}ab\\ \mu \end{subarray}} = \sum_{\sigma \nu}\sum_{cd}\Gamma_{\begin{subarray}{c} ab; cd \\ 0, \mu ; \sigma, \nu\end{subarray}} \rho_{\begin{subarray}{c}cd\\ \sigma \nu \end{subarray}} \\
&= \sum_{\sigma \nu}\sum_{cd}\Gamma_{\begin{subarray}{c} ab; cd \\ 0, \mu ; \sigma, \nu\end{subarray}} \rho_{\begin{subarray}{c}cd\\ \nu -\sigma \end{subarray}} =
\sum_{\nu}\sum_{cd} \left[ \sum\limits_{\sigma} \Gamma_{\begin{subarray}{c} ab; cd \\ 0, \mu ; \sigma, \sigma+\nu \end{subarray}} \right] \rho_{\begin{subarray}{c}cd\\ \nu \end{subarray}} \\
\therefore & \Gamma_{\begin{subarray}{c} ab; cd \\ \mu ; \nu \end{subarray}} = \left[ \sum\limits_{\sigma} \Gamma_{\begin{subarray}{c} ab; cd \\ 0, \mu ; \sigma, \sigma+\nu \end{subarray}} \right]
.
\end{split}
\end{equation}
Writing out this form explicitly for subbands and in-plane energy and taking advantage of the fact that
\begin{equation}
V_{\begin{subarray}{c} ab \\ \mu, \nu \end{subarray}} V_{\begin{subarray}{c} cd \\ \sigma, \eta \end{subarray}} = V_{\begin{subarray}{c} ab \\ \mu+\gamma, \nu+\gamma \end{subarray}} V_{\begin{subarray}{c} cd \\ \sigma+\gamma, \eta+\gamma \end{subarray}}
\end{equation} for any $\gamma$ leads to Eq. \ref{eq:gamma_compact_eres_period}. 

The incoherent velocity operator\cite{burnett_design_2016} can be inferred from
\begin{equation}
v_{ab} = \sum\limits_{cd} \Gamma^*_{cd,ab}z_{cd}.
\end{equation}
Taking advantage of the periodic submatrix form of $v$, we can write
\begin{equation}
\begin{split}
v_{\begin{subarray}{c}ab\\ 0 \mu \end{subarray}} &= v_{\begin{subarray}{c}ab\\ \mu \end{subarray}} = \sum_{\sigma \nu}\sum_{cd}\Gamma_{\begin{subarray}{c} cd ; ab\\ \sigma, \nu; 0, \mu \end{subarray}} z_{\begin{subarray}{c}cd\\ \sigma \nu \end{subarray}} \\
&= \sum_{\sigma \nu}\sum_{cd}\Gamma_{\begin{subarray}{c} cd ; ab \\ \sigma, \nu ; 0, \mu \end{subarray}} (z_{\begin{subarray}{c}cd\\ \nu -\sigma \end{subarray}} + \delta_{rs}\delta_{cd}\sigma L) \\
&= \sum_{\nu}\sum_{cd} \left[ \sum\limits_{\sigma} \Gamma_{\begin{subarray}{c} cd ; ab \\ \sigma, \sigma+\nu ; 0, \mu \end{subarray}} \left(1 + \frac{\delta_{rs}\delta_{cd}\sigma L}{z_{\begin{subarray}{c} cc \\ 0\end{subarray}}}\right) \right] z_{\begin{subarray}{c}cd\\ \nu \end{subarray}}
.
\end{split}
\end{equation}
As before, we can rewrite the term in brackets as a superoperator. Using labels for the subband and transverse momentum, we arrive at the incoherent scattering superoperator in Eq. \ref{eq:incohj_so}.

\section{Density Matrix Eigenstates and their Velocities}\label{app:dmeig}
We wish to show that an arbitrary two-level system, when rotated from the energy eigenstate to the DM eigenstate basis, will have equal and opposite coherent velocities and differing incoherent velocities for the DM eigenstates.
For a two-level system in the energy eigenstate representation, the density matrix in general will be
\begin{equation}
\rho = \begin{bmatrix}
\rho_A & \rho_{AB} \\ \rho_{AB}^* & \rho_B
\end{bmatrix}.
\end{equation}
Its eigenvalues are
$\rho_{1,2} = \dfrac{\rho_A + \rho_B \pm \Delta}{2}$ where $\Delta = \sqrt{(\rho_A - \rho_B)^2 + 4|\rho_{AB}|^2}$ and the indices $1, 2$ correspond to the plus and minus sign respectively, with $\rho_1$ being more populous. The eigenstates are
\begin{equation}
\begin{split}
\ket{\phi_{1}} = \begin{pmatrix}
\cos \theta_{1} \\ \sin \theta_{1}
\end{pmatrix} , \ket{\phi_{2}} = \begin{pmatrix}
\sin^*\theta_{1} \\ -\cos \theta_{1}
\end{pmatrix}
\end{split}
\end{equation}
where
\begin{equation}
\begin{split}
\cos \theta_1 &= \frac{\rho_A - \rho_B + \Delta}{\sqrt{4|\rho_{AB}|^2 + [\rho_A - \rho_B + \Delta]^2}} \\
\sin \theta_1 &= \frac{\rho_{AB}^*}{\sqrt{4|\rho_{AB}|^2 + [\rho_A - \rho_B + \Delta]^2}}
\end{split}
\end{equation}
Since the current is given by $\Tr(\rho v)$, for a diagonalized $\rho$ we are only interested in the diagonals of $v$. Upon transforming a general velocity operator $v$ from the energy eigenstate to DM eigenstate basis, we find
\begin{align*}
v_1 &= v_A \cos^2 \theta_1 + v_B |\sin \theta_1|^2 + 2 \cos \theta_1 \Re(v_{AB} \sin \theta_1) \\
v_2 &= v_A |\sin \theta_1|^2 + v_B \cos^2 \theta_1 - 2 \cos \theta_1 \Re(v_{AB} \sin \theta_1) .
\end{align*}
Since the coherent velocity operator in the energy basis only has a nonzero off-diagonal component ($v_{coh,A}=v_{coh,B}=0$ and $v_{coh,AB}=v_{coh,BA}^* = (E_A - E_B)z_{AB}/i\hbar$), we see that for a two-level system in the DM basis $v_1 = -v_2$. The incoherent velocity operator generally has nonzero diagonal elements in the energy basis and therefore does not have equal and opposite values in the DM basis. In the case of coherent Rabi oscillations of a pure state, $\rho_1 = 1$ and $\rho_2=0$, so that the instantaneous current is given by $v_1$. Otherwise, dephasing leads to $0<\rho_{1,2}<1$ and the net current is given by $\rho_1 v_1 + \rho_2 v_2$. Once additional states are introduced, $v_1 \neq -v_2$ in general since coherences with other states become important. However, for any $N$-level system, while the coherent velocities of the various DM eigenstate will differ, their sum will always remain zero since Tr$(v_{coh})=0$ is invariant under basis transformations.

\bibliography{prbqclmodel}
\end{document}